\documentclass[reprint,10pt]{revtex4-1}%reprint
\usepackage{braket}
\usepackage{mathtools}
\usepackage{adjustbox}
\usepackage[utf8]{inputenc}

\usepackage{csquotes}
\usepackage[dvipsnames]{xcolor}
\usepackage{color}
\usepackage{tikz}
\usepackage{graphics}
\usepackage{float}
\usetikzlibrary{shapes.geometric,backgrounds,positioning,shapes.geometric,decorations.markings,decorations.pathreplacing,arrows,knots,hobby,angles,quotes}
\usepackage{amsmath}
\usepackage{amssymb}
\usepackage{amsthm}
\usepackage{braket}
\usepackage{bbm}

\usepackage[draft]{todonotes} 
\usepackage{subcaption}
\DeclareMathOperator{\tr}{tr}
\DeclareMathOperator{\cnot}{\textsc{cnot}}
\DeclareMathOperator{\swap}{\textsc{swap}}
\DeclareMathOperator{\cswap}{\textsc{cswap}}
\DeclareMathOperator{\bigswap}{\textsc{bigswap}}

\usepackage{hyperref}

\usepackage{cleveref}
\crefname{section}{Section}{Sections}
\Crefname{section}{Section}{Sections}
\crefname{equation}{}{}
\Crefname{equation}{}{}
\crefname{figure}{Figure}{Figures}
\Crefname{figure}{Figure}{Figures}
\crefname{appendix}{Appendix}{Appendices}
\Crefname{appendix}{Appendix}{Appendices}

\definecolor{color1}{RGB}{94,60,153}
\definecolor{color2}{RGB}{230,97,1}
\definecolor{color3}{RGB}{253,184,99}
\usepackage[draft]{todonotes} 
\usepackage{subcaption}
\usepackage{caption} 

\begin{document}

\begin{abstract} 

\end{abstract}

\title{Efficient Learning for Deep Quantum Neural Networks} 
\author{Kerstin Beer}
\email{kerstin.beer@itp.uni-hannover.de}
\author{Dmytro Bondarenko}
\email{dimbond@live.com}
\author{Terry Farrelly}
\email{farreltc@tcd.ie}
\author{Tobias J.\ Osborne}
\author{Robert Salzmann}
\email{rals.salzmann@web.de}
\author{Ramona Wolf}
\email{ramona.wolf@itp.uni-hannover.de}

\affiliation{Institut f\"ur Theoretische Physik, Leibniz Universit\"at Hannover, Germany}
\maketitle

\textbf{Neural networks enjoy widespread success in both research and industry and, with the imminent advent of quantum technology, it is now a crucial challenge to design quantum neural networks for fully quantum learning tasks. 
Here we propose the use of quantum neurons as a building block for quantum feed-forward neural networks capable of universal quantum computation.  We describe the efficient training of these networks using the fidelity as a cost function and provide both classical and efficient quantum implementations.  Our method allows for fast optimisation with reduced memory requirements:\ the number of qudits required scales with only the width, allowing the optimisation of deep networks. We benchmark our proposal for the quantum task of learning an unknown unitary and find  remarkable generalisation behaviour and a striking robustness to noisy training data.}

Machine learning (ML), particularly applied to deep neural networks via the backpropagation algorithm, has enabled a wide spectrum of revolutionary applications ranging from the social to the scientific \cite{Goodfellow2016,Nielsen2015}. Triumphs include the now everyday deployment of handwriting and speech recognition through to applications at the frontier of scientific research \cite{Nielsen2015,Jordan2015,Bishop2006}. Despite rapid theoretical and practical progress, ML training algorithms are computationally expensive and, now that Moore’s law is faltering, we must contemplate a future with a slower rate of advance \cite{Prati2017}. However, new exciting possibilities are opening up due to the imminent advent of quantum computing devices that directly exploit the laws of quantum mechanics to evade the technological and thermodynamical limits of classical computation \cite{Prati2017}.

The exploitation of quantum computing devices to carry out \emph{quantum maching learning} (QML) is in its initial exploratory stages \cite{Biamonte2017}. One can exploit classical ML to improve quantum tasks (``QC'' ML, see \cite{Wikipedia2018} for a discussion of this terminology) such as the simulation of many-body systems \cite{Carleo2017}, adaptive quantum computation \cite{Tiersch2015} or quantum metrology \cite{Lovett2013}, or one can exploit quantum algorithms to speed up classical ML (``CQ'' ML) \cite{Aimeur2013,Paparo2014,Schuld2014,Wiebe2016}, or, finally, one can exploit quantum computing devices to carry out learning tasks with quantum data (``QQ'' ML) \cite{Amin2018,Alvarez2017,Du2018}. Particularly relevant to the present work is the recent paper of Verdon, Pye, and Broughton \cite{Verdon2018} where quantum learning of parametrised unitary operations is carried out coherently. There are still many challenging open problems left for QML, particularly, the task of developing quantum algorithms for learning tasks involving \emph{quantum data}. 

A series of hurdles face the designer of a QML algorithm for quantum data. These include, finding the correct quantum generalisation of the perceptron,  (deep) neural network architecture, optimisation algorithm, and loss function. In this paper we meet these challenges and propose a natural quantum perceptron which, when integrated into a quantum neural network (QNN), is capable of carrying out universal quantum computation. Our QNN architecture allows for a quantum analogue of the classical backpropagation algorithm by exploiting completely positive layer transition maps. We apply our QNN to the task of learning an unknown unitary, both with and without errors. Our classical simulation results are very promising and suggest the imminent feasibility of our procedure for noisy intermediate scale (NISQ) quantum devices.

% here we describe of a (deep) quantum feed forward neural network
%
\begin{figure}
\begin{tikzpicture}[scale=1]
\draw[white](0,-2.2)--(8,-2.2);
\draw[decorate,decoration={brace,amplitude=3pt}] 
(.75,-2.5) node(t_k_unten){} -- 
(2.75,-2.5) node(t_k_opt_unten){}; 
\node at (1.75,-2.1){$U^1=U_3^1U_2^1U_1^1$};
\end{tikzpicture}
\includegraphics[width=0.95\linewidth]{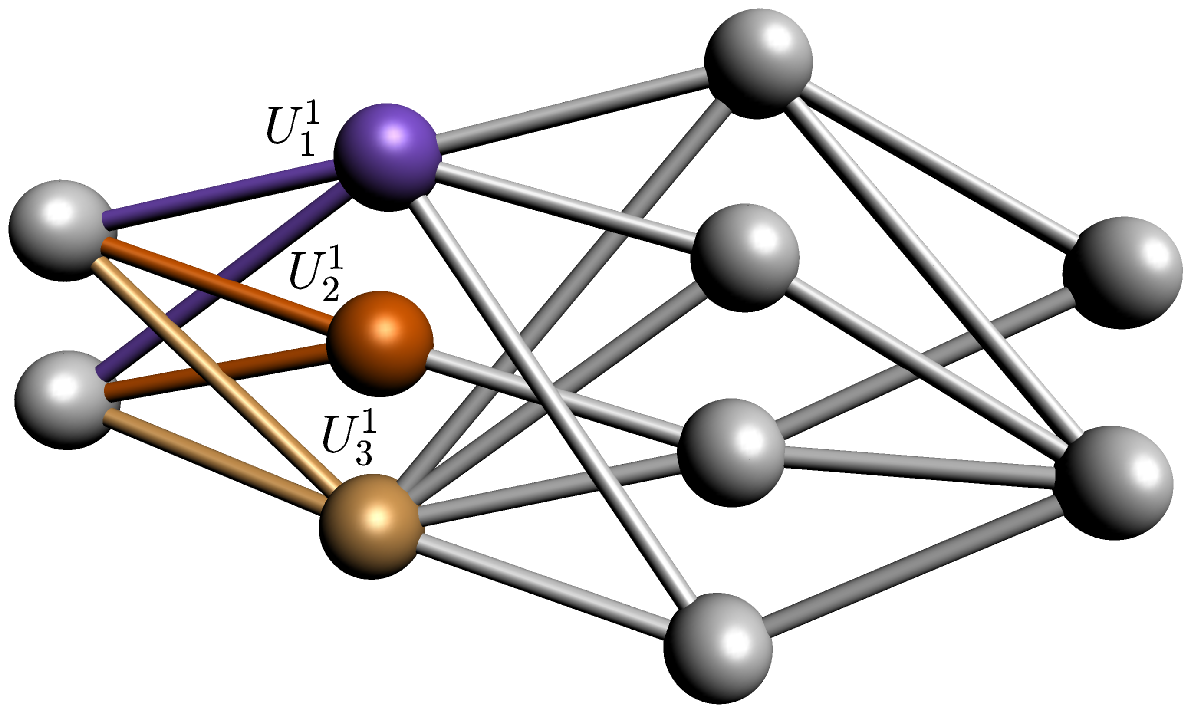}

\begin{tikzpicture}[scale=1]
\draw[decorate,decoration={brace,amplitude=3pt,mirror}] 
(7,-2.5) node(t_k_unten){} -- 
(8,-2.5) node(t_k_opt_unten){}; 
\node at (7.25,-3){output layer};
\node at (7.5,-2.25){};
\draw[decorate,decoration={brace,amplitude=3pt,mirror}] 
(0,-2.5) node(t_k_unten){} -- 
(.75,-2.5) node(t_k_opt_unten){}; 
\node at (.75,-3){input layer};
\node at (.4,-2.25){};
\draw[decorate,decoration={brace,amplitude=3pt,mirror}] 
(1,-2.5) node(t_k_unten){} -- 
(6.75,-2.5) node(t_k_opt_unten){}; 
\node at (2.5,-2.25){$l=1$};
\node at (4,-3){hidden layers};
\node at (3.75,-2.25){$\cdots$};
\node at (5,-2.25){$l=L$};
\end{tikzpicture}
\captionof{figure}{\textbf{A general quantum feed forward neural network.} A QNN has an \emph{input}, \emph{output}, and $L$ \emph{hidden} layers. We apply the perceptron unitaries layerwise from top to bottom (indicated with colours for the first layer): first the violet unitary is applied, followed by the orange one, and finally the yellow one.}\label{fig:QNN}
\end{figure}

There are now several available quantum generalisations of the \emph{perceptron}, the fundamental building block of a neural network \cite{Nielsen2015,Goodfellow2016,Schuld2015,lewenstein_quantum_1994,wan_quantum_2017,DASILVA201655,2001arxiv7012A,NQ1008,KoudaMNP05}.
In the context of CQ learning (in contrast to QQ learning, which we consider here) proposals include \cite{Torrontegui2018,FN18,SBS18,MNKF18}, who exploit a qubit circuit setup, though the gate choices and geometry are somewhat more specific than ours. Another interesting approach is to use continuous-variable quantum systems (e.g., light) to define quantum perceptrons \cite{KBA18,ABIMBK19,2018arXiv180810047S}.

With the aim of building a fully quantum deep neural network capable of universal quantum computation we have found it necessary to modify the extant proposals somewhat. In this paper we define a quantum perceptron to be a general unitary operator acting on $m$ input qubits and $n$ output qubits. The input qubits are initialised in a possibly unknown \emph{mixed} state $\rho^{\mathrm{in}}$ and the output qubits in a fiducial product state $|0\cdots0\rangle_{\text{out}}$. Our perceptron is then simply an \emph{arbitary} unitary applied to the $m+n$ input and output qubits. Such an arbitrary unitary operator depends on $(2^{m+n})^2-1$ parameters, which incorporate the weights and biases of previous proposals in a natural way (see the supplementary material for further details and the extension to qudits.) For simplicity in the sequel we focus on the case where our perceptrons act on $m$ input qubits and one output qubit, i.e., they are $(m+1)$-qubit unitaries.

Now we have a quantum neuron we can describe our quantum neural network architecture. Motivated by analogy with the classical case and consequent operational considerations (see the supplementary material for further details) we propose that a QNN is a quantum circuit of quantum perceptrons organised into $L$ \emph{hidden} layers of qubits, acting on an initial state $\rho^{\text{in}}$ of the \emph{input} qubits, and producing an, in general, mixed state $\rho^{\text{out}}$ for the \emph{output} qubits according to
\begin{equation}
	\rho^{\mathrm{out}} \equiv \tr_\mathrm{in,hid} \left( \mathcal{U}(\rho^{\mathrm{in}} \otimes |0\cdots0\rangle_{\text{hid,out}}\langle 0\cdots 0|) \mathcal{U}^\dag\right),
\end{equation}
where $\mathcal{U} \equiv U^\mathrm{out} U^L U^{L-1} \cdots U^1$ is the QNN quantum circuit, $U^l$ are the layer unitaries, comprised of a product of quantum perceptrons acting on the qubits in layers $l-1$ and $l$. It is important to note that, because our perceptrons are arbitrary unitary operators, they do not, in general, commute, so that the order of operations is significant. See Fig.~\ref{fig:QNN} for an illustration. 

It is a direct consequence of the quantum-circuit structure of our QNNs that they can carry out universal quantum computation. More remarkable, however, is the observation that a QNN comprised of quantum perceptrons acting on $4$-level qudits that commute \emph{within each layer}, is still capable of carrying out universal quantum computation (see the supplementary material for details). Although commuting qudit perceptrons suffice, we have actually found it convenient in practice to exploit noncommuting perceptrons acting on qubits.  In fact, the most general form of our quantum perceptrons can implement any quantum channel on the input qudits (see the supplemental material), so one could not hope for any more general notion of a quantum perceptron.

A crucial property of our QNN definition is that the network output may be expressed as the \emph{composition} of a sequence of \emph{completely positive} layer-to-layer transition maps $\mathcal{E}^l$:
\begin{equation}\label{eq:CPmapform}
	\rho^{\text{out}}= \mathcal{E}^\mathrm{out}\left(\mathcal{E}^{L}\left(\dots \mathcal{E}^{2}\left(\mathcal{E}^{1}\left(\rho^{\text{in}}\right)\right)\dots\right)\right),
\end{equation}
where $\mathcal{E}^{l}(X^{l-1}) \equiv \text{tr}_{l-1}\big(\prod_{j=m_l}^{1} U^l_j (X^{l-1}\otimes |0\cdots 0\rangle_l \langle 0\cdots 0|)$ $\prod_{j=1}^{m_l} {U_j^l}^\dag\big)$, $U_j^l$ is the $j$th perceptron acting on layers $l-1$ and $l$, and $m_l$ is the total number of perceptrons acting on layers $l-1$ and $l$. This characterisation of the output of a QNN highlights a key structural characteristic: information propagates from input to output and hence naturally implements a quantum \emph{feed-forward} neural network.
This key result is the fundamental basis for our quantum analogue of the backpropagation algorithm. 

As an aside, we can justify our choice of quantum perceptron for our QNNs, by contrasting it with a recent notion of a quantum perceptron as a \emph{controlled unitary} \cite{CGA17,Torrontegui2018}, i.e., $U = \sum_{\alpha} |\alpha\rangle\langle\alpha|\otimes U(\alpha)$, where $|\alpha\rangle$ is some basis for the input space and $U(\alpha)$ are parametrised unitaries. Substituting this definition into (\ref{eq:CPmapform}) implies that the output state is the result of a measure-and-prepare, or \emph{cq}, channel. That is, $\rho^{\text{out}} = \sum_{\alpha} \langle\alpha|\rho^{\text{in}}|\alpha\rangle U(\alpha)|0\rangle\langle0|U(\alpha)^\dag$. Such channels have no nonzero quantum channel capacity and cannot carry out general quantum computation.

% here we describe the quantum backpropagation algorithm

Now that we have an architecture for our QNN we can specify the learning task. Here we focus on the scenario where we have repeatable access to training data in the form of pairs $\left(|\phi^{\text{in}}_x\rangle, |\phi^{\text{out}}_x\rangle\right)$, $x = 1,2,\ldots, N$, of possibly unknown quantum states. (It is crucial that we can request multiple copies of a training pair $\left(|\phi^{\text{in}}_x\rangle, |\phi^{\text{out}}_x\rangle\right)$ for a specified $x$ in order to overcome quantum projection noise in evaluating the derivative of the cost function.) For concreteness in the sequel we focus on the restricted case where $|\phi^{\text{out}}_x\rangle = V|\phi^{\text{in}}_x\rangle$, where $V$ is some unknown unitary operation. This scenario is typical when one has access to an untrusted or uncharacterised device which performs an unknown quantum information processing task and one is able to repeatably initialise and apply the device to arbitrary initial states.

To evaluate the performance of our QNN in learning the training data, i.e., how close is the network output $\rho^{\text{out}}_x$ for the input $|\phi^{\text{in}}_x\rangle$ to the correct output $|\phi^{\text{out}}_x\rangle$, we need a \emph{cost function}. Operationally, there is an essentially unique measure of closeness for (pure) quantum states, namely the fidelity, and it is for this reason that we define our cost function to be the fidelity between the QNN output and the desired output averaged over the training data:
\begin{equation}
	C=\frac{1}{N}\sum_{x=1}^N \langle\phi^{\text{out}}_x\rvert\rho_x^{\text{out}}\lvert\phi^{\text{out}}_x\rangle.
\end{equation}
Note that the cost function takes a slightly more complicated form when the training data output states are not pure \footnote{In that case, we simply use the fidelity for mixed states:\ $F(\rho,\sigma):=\left[\mathrm{tr}\sqrt{\rho^{1/2}\sigma\rho^{1/2}}\right]^2$.}, which may occur if we were to train our network to learn a quantum channel.
The cost function varies between $0$ (worst) and $1$ (best).
\begin{figure*}
	\begin{minipage}{.48\textwidth}
\scriptsize
\flushleft	
\textbf{1. Initialize:}\\
Choose the initial $U_j^l$ randomly for all $j$ and $l$.\\
\hspace{2pt}\\
\textbf{2. Feedforward:} 
For every training pair $\left(\ket{\phi^\mathrm{in}_x},\ket{\phi^\mathrm{out}_x}\right)$ and every layer $l$, perform the following steps:\\
\textbf{2a.} Apply the channel $\mathcal{E}^l$ to the output state of layer $l-1$: Tensor $\rho_x^{l-1}$ with layer $l$ in state $\lvert 0\dots 0\rangle_l$ and apply $U^l=U_{m_l}^l\dots U_1^l$:
	\begin{figure}[H]
		\centering
		\includegraphics[width=0.97\linewidth]{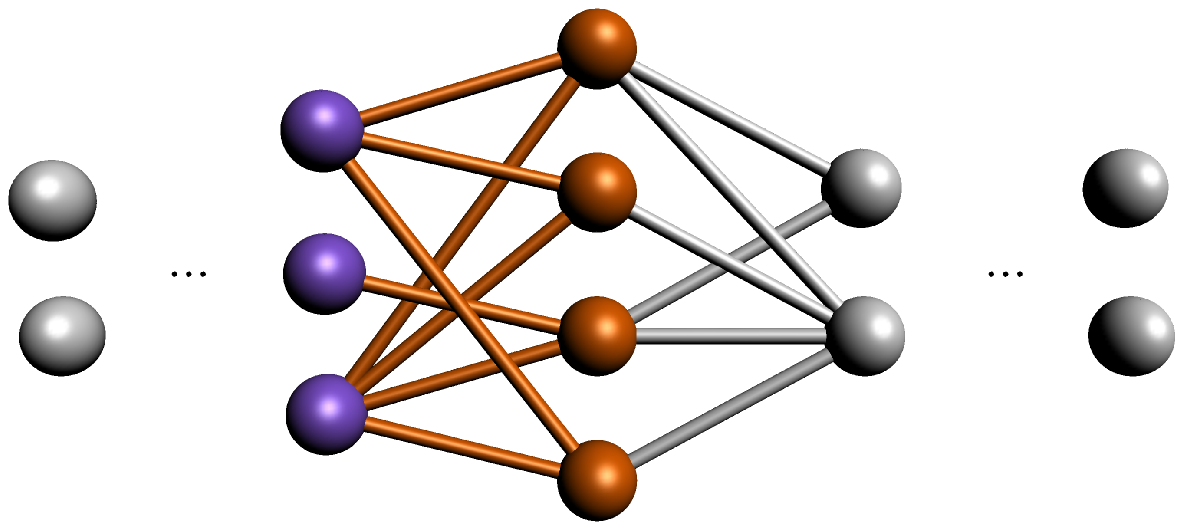}\\
		\begin{tikzpicture}[scale=1.07]
		\node at (7.5,-2.25){out};
		\node at (.3,-2.25){in};
		\node at (2,-2.25){$l-1$};
		\node at (4,-2.25){$l$};
		\node at (5.75,-2.25){$l+1$};
		\end{tikzpicture}
	\end{figure}
		\textbf{2b.} Trace out layer $l-1$ and store $\rho_x^l$.\\
\hspace{2pt}\\
\textbf{3. Update the network:} \\
\textbf{3a.} Calculate the parameter matrices given by
				\begin{equation*}
					K_j^l=\eta\frac{2^{m_{l-1}}}{N}\sum_{x=1}^N \mathrm{tr}_\mathrm{rest}M_j^l
				\end{equation*}

			where the trace is over all qubits that are not affected by $U_j^l$, $\eta$ is the learning rate and 
\end{minipage}
\hfill
\begin{minipage}{.48\textwidth}
\scriptsize
\flushleft	
				\begin{align*}
					M_j^l&=\Big[\prod_{\alpha=j}^1U_\alpha^l\left(\rho_x^{l-1,l}\right)\prod_{\alpha=1}^j{U_\alpha^l}^\dagger,\prod_{\alpha=j+1}^{m_l}{U_\alpha^l}^\dagger\left(\mathbb{I}_{l-1}\otimes\sigma_x^l\right)\prod_{\alpha=m_l}^{j+1}U_\alpha^l\Big],
				\end{align*}
	where $\rho_x^{l-1,l}=\rho_x^{l-1}\otimes\lvert0... 0\rangle_l\langle 0... 0\rvert$, $\sigma_x^l=\mathcal{F}^{l+1}\left(...\mathcal{F}^\mathrm{out}\left(\lvert\phi^\mathrm{out}_x\rangle\langle\phi_x^\mathrm{out}\rvert\right)...\right)$ and $\mathcal{F}^l$ is the adjoint channel to $\mathcal{E}^l$, i.e.\ the transition channel from layer $l+1$ to layer $l$. Below, the two parts of the commutator are depicted:
	\begin{figure}[H]
		\centering
		\begin{minipage}{.48\textwidth}
			\centering
\includegraphics[width=1\linewidth]{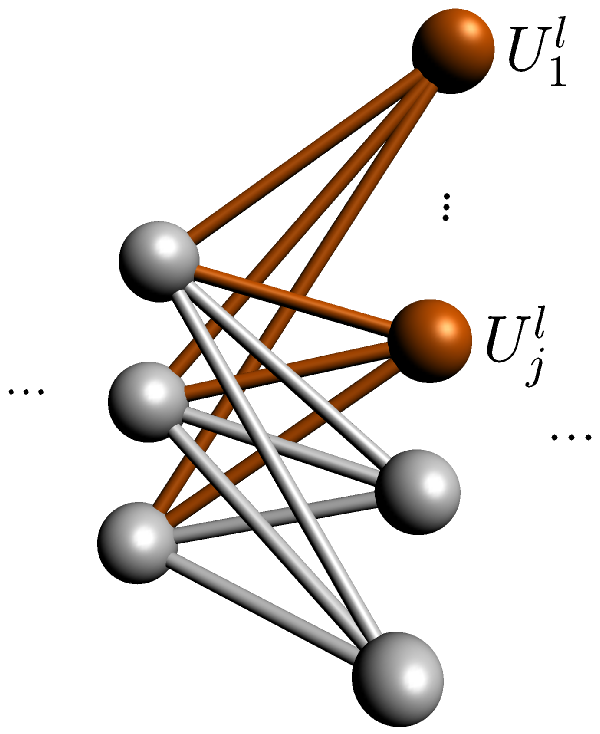}

\begin{tikzpicture}[scale=.6]
\draw[decorate,decoration={brace,amplitude=3pt,mirror}] 
(0,-2.5) node{} -- (1.75,-2.5) node{}; 
\node at (1,-3){$l-1$};
\draw[decorate,decoration={brace,amplitude=3pt,mirror}] 
(2.5,-2.5) {} -- (4.25,-2.5) {}; 
\node at (3.5,-3){$l$};
\draw[white,decorate,decoration={brace,amplitude=3pt}] 
(-1.75,-2.5) {} -- (2,-2.5) {}; 
\node at (2.5,-3.8){$\rho^{l-1}\otimes\ket{0\hdots 0}_l\bra{0\hdots 0}$};
\end{tikzpicture}
\end{minipage}
\begin{minipage}{.48\textwidth}
\includegraphics[width=1\textwidth]{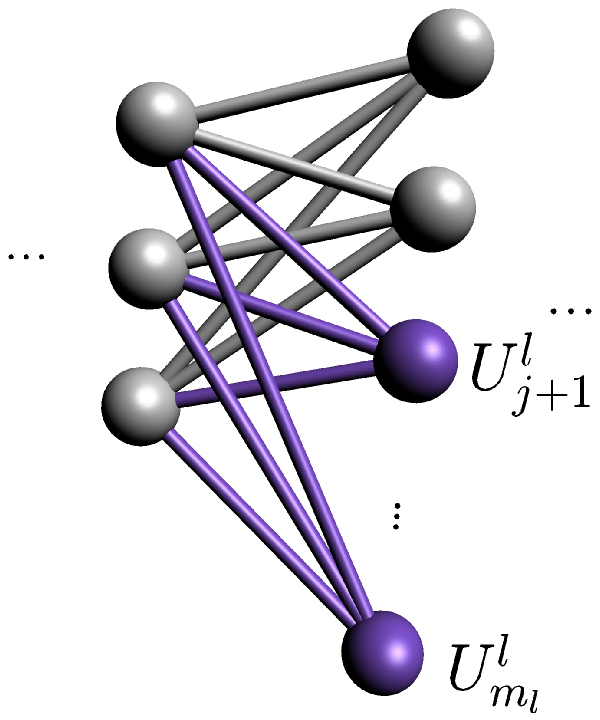}

	\begin{tikzpicture}[scale=.6]
\draw[decorate,decoration={brace,amplitude=3pt,mirror}] 
(0,-2.5) node{} -- (1.75,-2.5) node{}; 
\node at (1,-3){$l-1$};
\draw[decorate,decoration={brace,amplitude=3pt,mirror}] 
(2.5,-2.5) {} -- (4.25,-2.5) {}; 
\node at (3.5,-3){$l$};
\draw[white,decorate,decoration={brace,amplitude=3pt}] 
(-1.,-2.5) {} -- (2,-2.5) {}; 
\node at (2.5,-3.8){$\mathbbm{1}_{l-1}\otimes\sigma^l$};
\end{tikzpicture}	
\end{minipage}
	\end{figure}
\textbf{3b.} Update each unitary $U_j^l$ according to $U_j^l\rightarrow e^{i\epsilon K_j^l} U_j^l$.\\
\hspace{2pt}\\
\textbf{4. Repeat:} Repeat step 2. and 3. until the cost function reaches its maximum.
\end{minipage}

\captionof{figure}{\textbf{Training algorithm.}}
\label{fig:alg}
\end{figure*}

We train the QNN by optimising the cost function $C$. This, as in the classical case, proceeds via update of the QNN parameters: at each training step, we update the perceptron unitaries according to $U\rightarrow e^{i\epsilon K}U$, where $K$ is the matrix that includes all parameters of the corresponding perceptron unitary and $\epsilon$ is the chosen step size. The matrices $K$ are chosen so that the cost function increases most rapidly: the change in $C$ is given by
	\begin{equation}
	\label{eq:deltaC}
		\Delta C = \frac{\epsilon}{N}\sum_{x=1}^{N}\sum_{l=1}^{L+1} \text{tr}\left(\sigma^l_x \Delta\mathcal{E}^{l}\left(\rho^{l-1}_x\right)\right),
	\end{equation}
where $L+1=\mathrm{out}$, $\rho^l_x= \mathcal{E}^{l}\left(\cdots \mathcal{E}^{2}\left(\mathcal{E}^{1}\left(\rho^{\text{in}}_x\right)\right)\dots\right)$, $\sigma^l_x=\mathcal{F}^{l+1}\left(\cdots \mathcal{F}^{L}\left(\mathcal{F}^\mathrm{out}\left(\lvert\phi_x^\mathrm{out}\rangle\langle\phi_x^\mathrm{out}\rvert\right)\right)\cdots\right)$, and $\mathcal{F}(X) \equiv \sum_{\alpha} A_\alpha^\dag X A_\alpha$ is the \emph{adjoint channel} for the CP map $\mathcal{E}(X)=\sum_{\alpha} A_\alpha X A_\alpha^\dag$. From \eqref{eq:deltaC}, we obtain a formula for the parameter matrices (this is described in detail in the supplementary material). At this point, the layer structure of the network comes in handy: To evaluate $K_j^l$ for a specific perceptron, we only need the output state of the previous layer, $\rho^{l-1}$ (which is obtained by applying the layer-to-layer channels $\mathcal{E}^1,\mathcal{E}^2\dots\mathcal{E}^{l-1}$ to the input state), and the state of the following layer $\sigma^l$ obtained from applying the adjoint channels to the desired output state up to the current layer (see Fig.~\ref{fig:alg}). A striking feature of this algorithm is that the parameter matrices may be calculated layer-by-layer without ever having to apply the unitary corresponding to the full quantum circuit on all the constituent qubits of the QNN in one go.  In other words, we need only access two layers at any given time, which greatly reduces the memory requirements of the algorithm.  Hence, the size of the matrices in our calculation only scale with the width of the network, enabling us to train deep QNNs.

\begin{figure*}
\begin{subfigure}{0.49\textwidth}

\begin{tikzpicture}[yscale=5,xscale=1]
\def\xmin{0}
\def\xmax{8}
\def\ymin{0}
\def\ymax{1}
\draw[->, thin, draw=gray] (\xmin,\ymin)--(\xmax,\ymin) node [below left] {\tiny Number of training pairs}; 
\draw[->, thin, draw=gray] (\xmin,\ymin)--(\xmin,\ymax)node [above right] {\tiny Cost for test pairs};  
\foreach \x in {1,...,5}
\node at (\x, \ymin) [below] {\tiny\x};
\foreach \y in {.1,.2,.3,.4,.5,.6,.7,.8,.9}
\node at (\xmin,\y) [left] {\tiny\y};
%put in the next line data after "in"
\foreach \Point in {(1, 0.22019666130478677),(2, 0.34456619098462443),(3, 0.4672191059180644),(4, 0.6153667918789689),(5, 0.7393789506540993),(6, 0.8569043801723268),(7, 0.9481239799787362),(8, 0.9854768291157662)}{
\node[color2] at \Point {\textbullet};
}
\foreach \Point in {(1, 0.225), (2, 0.344444), (3, 0.475), (4, 0.608333), (5, 0.736111), (6, 0.85), (7, 0.941667), (8, 1)}{
\node[color1] at \Point {\textbullet};
}
\draw [xstep=1,ystep=.1,gray, dotted]  (8,0) grid (0,1);
\end{tikzpicture}
\caption{\textbf{Generalisation.} We trained the network
	\begin{tikzpicture}[scale=0.65, baseline]
		\node(1) [circle,draw,inner sep=0pt,minimum size=4.5pt] at (-1,0) {};
		\node(2) [circle,draw,inner sep=0pt,minimum size=4.5pt] at (-1,0.3) {};
		\node(3) [circle,draw,inner sep=0pt,minimum size=4.5pt] at (-1,0.6) {};
		\node(4) [circle,draw,inner sep=0pt,minimum size=4.5pt] at (0,0) {};
		\node(5) [circle,draw,inner sep=0pt,minimum size=4.5pt] at (0,0.3) {};
		\node(6) [circle,draw,inner sep=0pt,minimum size=4.5pt] at (0,0.6) {};
		\node(7) [circle,draw,inner sep=0pt,minimum size=4.5pt] at (1,0) {};
		\node(8) [circle,draw,inner sep=0pt,minimum size=4.5pt] at (1,0.3) {};
		\node(9) [circle,draw,inner sep=0pt,minimum size=4.5pt] at (1,0.6) {};
		\draw (1)--(4);
		\draw (2)--(4);
		\draw (3)--(4);
		\draw (1)--(5);
		\draw (2)--(5);
		\draw (3)--(5);
		\draw (1)--(6);
		\draw (2)--(6);
		\draw (3)--(6);
		\draw (4)--(7);
		\draw (5)--(7);
		\draw (6)--(7);
		\draw (4)--(8);
		\draw (5)--(8);
		\draw (6)--(8);
		\draw (4)--(9);
		\draw (5)--(9);
		\draw (6)--(9);
\end{tikzpicture}
	(unitaries applied from top to bottom)
	 with $\epsilon=0.1$, $\eta=2/3$ for $1000$ rounds with $n=1,2,\dots,8$ training pairs and evaluated the cost function for a set of $10$ test pairs afterwards. We averaged this over $20$ rounds (orange points) and compared the result to the estimated value of the optimal achievable cost function (violet points).}
 \label{fig:resultsa}
\end{subfigure}
\hfill
\begin{subfigure}{0.49\textwidth}
\raisebox{3.9mm}{
\begin{tikzpicture}[yscale=5,xscale=.08]
\def\xmin{0}
\def\xmax{100}
\def\ymin{0}
\def\ymax{1}
\draw[->, thin, draw=gray] (\xmin,\ymin)--(\xmax,\ymin) node [below left] {\tiny Number of noisy pairs}; 
\draw[->, thin, draw=gray] (\xmin,\ymin)--(\xmin,\ymax) node [above right] {\tiny Cost for good test pairs};  
\foreach \x in {10,20,...,60}
\node at (\x, \ymin) [below] {\tiny\x};
\foreach \y in {.1,.2,.3,.4,.5,.6,.7,.8,.9}
\node at (\xmin,\y) [left] {\tiny\y};
%put in the next line data after "in"
\foreach \Point in {(0, 0.9999988812037631),(5, 0.9998675362562289),(10, 0.9991545815565356),(15, 0.9980864880626191),(20, 0.9981773438490551),(25, 0.9973477437970019),(30, 0.996208256972089),(35, 0.9948700240610997),(40, 0.9947395171844108),(45, 0.9924308541328096),(50, 0.9865230944649602),(55, 0.9883202771721912),(60, 0.9733380112449577),(65, 0.9607046692132285),(70, 0.9579410995168592),(75, 0.8729391963495594),(80, 0.7336290899974561),(85, 0.5779847247543635),(90, 0.4171436560766765),(95, 0.2559509395527556),(100, 0.23817485576276787)}{
\node[color2] at \Point {\textbullet};
}
\draw [xstep=10,ystep=.1,gray, dotted]  (100,0) grid (0,1);
\end{tikzpicture}}
\caption{\textbf{Robustness of the QNN to noisy data.} We trained the network
	\begin{tikzpicture}[scale=0.65, baseline]
	\node(11) [circle,draw,inner sep=0pt,minimum size=4.5pt] at (0,0) {};
	\node(12) [circle,draw,inner sep=0pt,minimum size=4.5pt] at (0,0.3) {};
	\node(21) [circle,draw,inner sep=0pt,minimum size=4.5pt] at (0.7,-0.15) {};
	\node(22) [circle,draw,inner sep=0pt,minimum size=4.5pt] at (0.7,0.15) {};
	\node(23) [circle,draw,inner sep=0pt,minimum size=4.5pt] at (0.7,0.45) {};
	\node(31) [circle,draw,inner sep=0pt,minimum size=4.5pt] at (1.4,0) {};
	\node(32) [circle,draw,inner sep=0pt,minimum size=4.5pt] at (1.4,0.3) {};
	\draw (11)--(21);
	\draw (12)--(21);
	\draw (11)--(22);
	\draw (12)--(22);
	\draw (11)--(23);
	\draw (12)--(23);
	\draw (21)--(31);
	\draw (22)--(31);
	\draw (23)--(31);
	\draw (21)--(32);
	\draw (22)--(32);
	\draw (23)--(32);
	\end{tikzpicture}
with $\epsilon=0.1$, $\eta=1$ for $300$ rounds with $100$ training pairs. In the plot, the number on the $x$-axis indicates how many of these pairs were replaced by a pair of noisy (i.e.\ random) pairs and the cost function is evaluated for all ``good'' test pairs.}
\label{fig:resultsb}

\end{subfigure}
\caption{\textbf{Numerical results.}}
\label{fig:results}
\end{figure*}
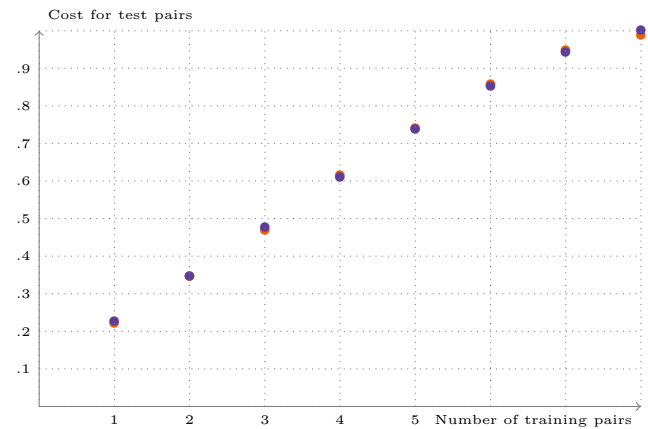
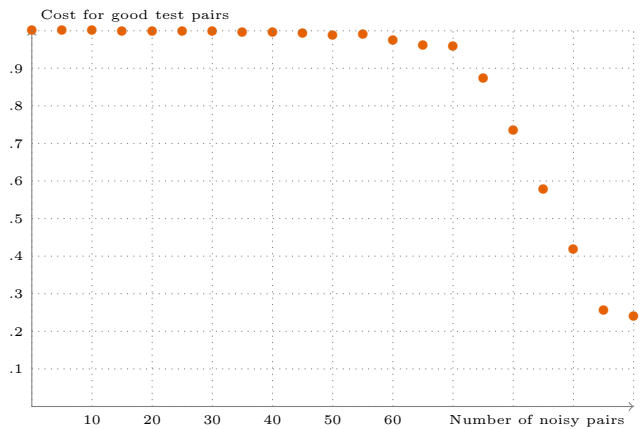
It is impossible to classically simulate deep QNN learning algorithms for more than a handful of qubits due to the exponential growth of Hilbert space. To evaluate the benchmark the performance of our QML algorithm we have thus been restricted to QNNs with small widths. We have carried out pilot simulations for input and output spaces of $m=2$ and $3$ qubits and have explored the behaviour of the QML gradient descent algorithm for the task of learning a random unitary $V$ (see supplementary material for the implementation details). We focussed on two separate tasks:\ In the first task we studied the ability of a QNN to generalise from a limited set of random training pairs $(|\phi_x^{\text{in}}\rangle, V|\phi_x^{\text{in}}\rangle)$, with $x=1,\ldots, N$, where $N$ was smaller than the Hilbert space dimension. The results are displayed in Fig~\ref{fig:resultsa}. Here we have plotted the (numerically obtained) cost function after training alongside a theoretical estimate of the optimal cost function for the best unitary possible which exploits all the available information (for which $C\sim \frac{n}{N}+\frac{N-n}{ND(D+1)}\left(D+\min\{n^2+1,D^2\}\right)$, where $n$ is the number of training pairs, $N$ the number of test pairs and $D$ the Hilbert space dimensions). Here we see that the QNN matches the theoretical estimate and demonstrates the remarkable ability of our QNNs to generalise. 

The second task we studied was aimed at understanding the robustness of the QNN to corrupted training data (e.g., due to decoherence). To evaluate this we generated a set of $N$ good training pairs and then corrupted $n$ of them by replacing them with random quantum data, where we chose the subset that was replaced by corrupted data randomly each time. We evaluated the cost function for the good pairs to check how well the network has learned the actual unitary. As illustrated in Fig~\ref{fig:resultsb} the QNN is extraordinarily robust to this kind of error.

A crucial consequence of our numerical investigates was the absence of a ``barren plateu'' in the cost function landscape for our QNNs \cite{McClean2018}. We always initialised our QNNs with random unitaries and we did not observe any exponential reduction in the value of the parameter matrices $K$ (which arise from the derivative of our QNN with respect to the parameters). This may be intuitively understood as a consequence of the nongeneric structure of our QNNs: at each layer we introduce new clean ancilla, which lead to an, in general, dissipative output.    

The QNN and training algorithm we have presented here lend themselves well to the coming era of NISQ devices. The network architecture enables a reduction in the number of coherent qubits required to store the intermediate states needed to evaluate a QNN. Thus we only need to store a number of qubits scaling with the width of the network. This remarkable reduction does come at a price, namely, we require multiple evaluations of the network to estimate the derivative of the cost function. However, in the near term, this tradeoff is a happy one as many NISQ architectures -- most notably superconducting qubit devices -- can easily and rapidly repeat executions of a quantum circuit. It is the task of adding coherent qubits that will likely be the challenging one in the near term and working with this constraint is the main goal here.

% here we discuss results and conclusions 

In this paper we have introduced natural quantum generalisations of perceptrons and (deep) neural networks, and proposed an efficient quantum training algorithm. The resulting QML algorithm, when applied to our QNNs, demostrates remarkable capabilities, including, the ability to generalise, tolerance to noisy training data, and an absence of a barren plateau in the cost function landscape. There are many natural questions remaining in the study of QNNs including: generalising the quantum perceptron definition further to cover general CP maps (thus incorporating a better model for decoherence processes), studying the effects of overfitting, and optimised implementation on the next generation of NISQ devices.

\begin{acknowledgments}
This work was supported by the DFG through SFB 1227 (DQ-mat), the RTG 1991, and Quantum Frontiers. Helpful correspondence and discussions with Lorenzo Cardarelli, Polina Feldmann, Alexander Hahn, Amit Jamadagni, Maria Kalabakov, Sebastian Kinnewig, Roger Melko, Laura Niermann, Simone Pfau, Deniz E.\ Stiegemann, and E.\ Miles Stoudenmire are gratefully acknowledged.
\end{acknowledgments}

\newpage
\onecolumngrid
\appendix
\section{A summary of the existing approaches for quantum perceptrons and quantum neural networks}

\subsection{Quantum algorithms for classical data}
	There are several proposals for efficiently training classical neural networks via quantum algorithms. In \cite{AHKZ18}, for example, the authors use a quantum subroutine to for efficiently approximating the inner products between vectors and store intermediate values in quantum random access memory. This yields a quadratically faster running time of their algorithm compared to the classical counterparts.
	
	Another interesting approach for learning classical data via quantum algorithms is using a qubit-circuit setup, which was done in \cite{FN18,SBS18,MNKF18,Gy18}, for example. Although the setup reminds of a quantum neural network, the gate choices and geometry differ from ours.
	
	One alternative option for quantum perceptrons and feedforward neural networks involves continuous-variable quantum systems \cite{KBA18,ABIMBK19}.  These are universal for continuous-variable quantum computation by virtue of including a non-Gaussian gate, and they are well-suited to CQ learning, as classical machine learning typically involves vectors in $\mathbb{R}^d$.
	
\subsection{Controlled unitaries as perceptrons}
	There have been several attempts to define a quantum analogue of the classical perceptron, of which many have used what we call the ``controlled unitary form'', i.e.\ unitaries of the form $U=\sum_\alpha\lvert\alpha\rangle\langle\alpha\rvert\otimes U(\alpha)$. In \cite{Torrontegui2018}, for example, the authors proposed perceptrons of the following form: the $j^{th}$ perceptron in the $l^{th}$ layer of the network is defined as a qubit with the following unitary acting on it:
	\begin{equation}
		\label{eq:qperceptron}
		\hat{U}_j^l\left(\hat{z}_j^l;f\right)=\exp\left[i \tilde{f}(\hat{z}_j^l)\ \hat{\sigma}_{j,l}^y\right],
	\end{equation}
	where $\tilde{f}(x)=\arcsin(f(x)^{\frac{1}{2}})$ with the activation function $f(\hat{z}_j^l):\mathbb{R}\to[0,1]$ and $\hat{z}_j^l=\sum_k w_{jk}^l \hat{\sigma}^z_{k,l-1}+b_j^l\mathbb{I}$. Note that within one layer, all unitaries commute.
	
	It is straightforward to see that these candidate perceptrons are not general enough for our purposes.  As we will see, they cannot create entanglement in the output state (meaning, for example, that they cannot be universal for quantum computing, though the authors of \cite{Torrontegui2018} never claimed this).  It is sufficient to look at the state of the $l$th layer after applying $U^l=\prod_{j}\hat{U}_j^l\left(\hat{z}_j^l;f\right)$, where the product is over all perceptron unitaries acting in the $l$th layer.  Notice that $U^l$ has the form
	\begin{equation}
	 U^l=\sum_{r_1,\dots,r_{m_{l-1}}\in\{0,1\}}\ket{r_1,\dots,r_{m_{l-1}}}_{l-1}\bra{r_1,\dots,r_{m_{l-1}}}\otimes \prod_{j=1}^{m_l}V^l_j(r_1,\dots,r_{m_{l-1}}),
	\end{equation}
where $\ket{r_1,\dots,r_{m_{l-1}}}_{l-1}$ is the state of the qubits in layer $l-1$ in the computational basis, $m_l$ is the number of qubits in the $l$th layer, and $V^l_j(r_1,...,r_{m_{l-1}})$ is a unitary that acts non-trivially only on qubit $j$ in layer $l$.  Suppose the state of these two layers before applying $U^l$ is $\rho_\mathrm{in}^{l-1}\otimes \rho_\mathrm{in}^{l}$ (this is actually more general than what is considered in \cite{Torrontegui2018}).  Then the state of layer $l$ after applying $U^l$ is given by
\begin{equation}
 \begin{split}
  \rho^l_\mathrm{out} & =\mathrm{tr}_{l-1}\left(U^l\rho_\mathrm{in}^{l-1}\otimes \rho_\mathrm{in}^{l}U^l\right)\\
  & =\sum_{r_1,\dots,r_{m_{l-1}}\in\{0,1\}}\bra{r_1,\dots,r_{m_{l-1}}}\rho_{\mathrm{in}}^{l-1}\ket{r_1,\dots,r_{m_{l-1}}}
  \prod_{j=1}^{m_l}V^l_j(r_1,\dots,r_{m_{l-1}})\ \rho_{\mathrm{in}}^{l}\ \prod_{j=1}^{m_l}{V_j^l}^\dagger(r_1,\dots,r_{m_{l-1}})\\
  & =\sum_{r_1,\dots,r_{m_{l-1}}\in\{0,1\}}p(r_1,\dots,r_{m_{l-1}})
  \prod_{j=1}^{m_{l}}V^l_j(r_1,\dots,r_{m_{l-1}})\ \rho_{\mathrm{in}}^{l}\ \prod_{j=1}^{m_l}{V_j^l}^\dagger(r_1,\dots,r_{m_{l-1}}),
 \end{split}
\end{equation}
where $p(r_1,\dots,r_{m_{l-1}})$ is a normalized probability distribution.  The crucial point is that, if $\rho_{\mathrm{in}}^{l}$ is a separable state of the qubits in layer $l$ (it is usually taken to be $\ket{0\dots 0}$), then the output state $\rho_{\mathrm{out}}^l$ is separable.  This is true regardless of whether the state of layer $l-1$ was entangled. Therefore, the output of each subsequent layer of the neural network is not entangled, which also applies to the output layer of the neural network.

\subsection{Implementation on near-term quantum computers}

With the first small quantum computers available, many people have studied how quantum machine learning (and quantum-assisted ML) proposals can be implemented on near-term quantum computing devices \cite{Biamonte2017,PBRB18,LJS19}.

Furthermore, there has been quite a lot of progress on programming languages and frameworks customized for the implementation of quantum ML tasks. Examples for this are, amongst others, the Python library PennyLane \cite{PennyLane}, that provides an architecture for ML of quantum and hybrid quantum-classical computations on near-term quantum-computing devices, and Strawberry Fields \cite{StrawberryFields}, a quantum programming architecture for light-based quantum computing (also built in Python), which allows for quantum ML of continuous-variable circuits.

% !TeX spellcheck = en_GB
	 
\section{The Quantum Neural Network}

	In this section, we describe a generalised version of the quantum perceptron. In contrast to extant proposals on quantum perceptrons, we define a quantum perceptron to be a \emph{general} unitary operator $U$ that is acting on $m$ input qudits and $n$ output qudits, where the input qudits are in a (possibly unknown) mixed state $\rho^\mathrm{in}$ and the output qudits in the product state $\lvert 0\dots 0\rangle$. The output of one layer of perceptrons is then 
		\begin{equation}
		\label{eq:Qperceptron}
			\rho^\mathrm{out}=\mathrm{tr}_\mathrm{in}\left(U_\mathrm{in,out}\left(\rho^\mathrm{in}\otimes\lvert 0\dots 0\rangle_\mathrm{out}\langle 0\dots 0\rvert\right)U_\mathrm{in,out}^\dagger\right),
		\end{equation}
	where $U_\mathrm{in,out}$ is the product of all unitaries in that layer.
	For simplicity (especially in the implementation), we focus on the case where our perceptrons act on $m$ input qubits and one output qubit. 
	
	This definition is motivated by the most general quantization of the classical machine learning scenario. In the typical machine learning framework (see e.g.\ \cite{Wolf2018}), the training data is a set of instances of some (unknown) probability distribution. And the natural quantisation of probability distributions are density matrices. Furthermore, the most general physical operations on density matrices are completely positive (CP, see e.g.\ \cite{wolf2012quantum} or \cite{2015arXiv150503106B}) maps. Due to the Stinespring dilation, the most general CP map can be written as in equation (\ref{eq:Qperceptron}), something which is expanded upon in more detail in section \ref{sec:universality}. In short, asking for the most general quantum version of a classical perceptron gives rise to the form we use here.
	
\begin{figure}[H]
	\centering
\begin{tikzpicture}[scale=1.15]
\draw[white](0,-2.2)--(8,-2.2);
\draw[decorate,decoration={brace,amplitude=3pt}] 
(.75,-2.5) node(t_k_unten){} -- 
(2.75,-2.5) node(t_k_opt_unten){}; 
\node at (1.75,-2.1){$U^1=U_3^1U_2^1U_1^1$};
\end{tikzpicture}

\includegraphics[width=.5\linewidth]{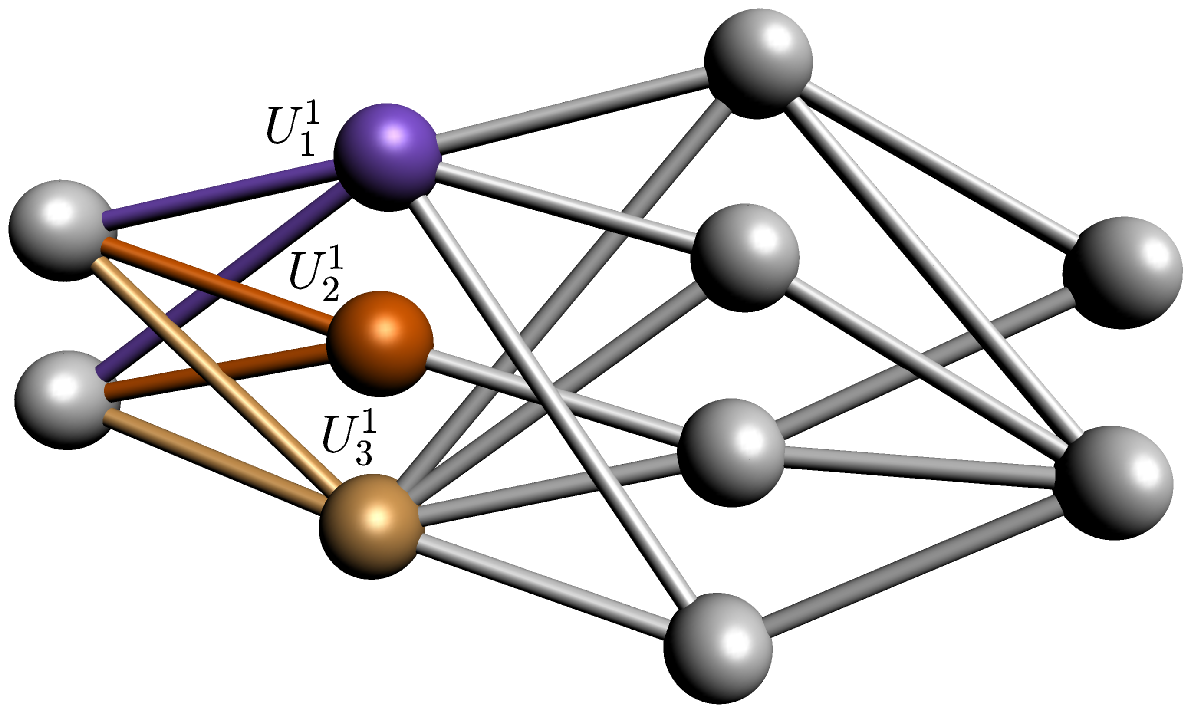}
	
\begin{tikzpicture}[scale=1.15]
\draw[decorate,decoration={brace,amplitude=3pt,mirror}] 
(7,-2.5) node(t_k_unten){} -- 
(8,-2.5) node(t_k_opt_unten){}; 
\node at (7.25,-3){output layer};
\node at (7.5,-2.25){$l=\text{out}$};
\draw[decorate,decoration={brace,amplitude=3pt,mirror}] 
(0,-2.5) node(t_k_unten){} -- 
(.75,-2.5) node(t_k_opt_unten){}; 
\node at (.75,-3){input layer};
\node at (.4,-2.25){$l=\text{in}$};
\draw[decorate,decoration={brace,amplitude=3pt,mirror}] 
(1,-2.5) node(t_k_unten){} -- 
(6.75,-2.5) node(t_k_opt_unten){}; 
\node at (2.5,-2.25){$l=1$};
\node at (4,-3){hidden layers};
\node at (3.75,-2.25){$\cdots$};
\node at (5,-2.25){$l=L$};
\end{tikzpicture}
\caption{A quantum feed forward neural network that has an \emph{input}, an \emph{output}, and $L$ \emph{hidden} layers. 
We marked the perceptron unitaries in the first layer with colors to demonstrate the order of the operations. We always apply the unitaries from top to bottom: first the violet one, followed by the orange one and the yellow one afterwards.}
\label{fig:genQNN}
\end{figure}

	Having a quantum perceptron in hand, we can now focus on the architecture of a full quantum neural network. As depicted in Figure~\ref{fig:genQNN}, a quantum neural network is built similarly to its classical counterpart: it consists of several layers, namely $L$ hidden layers and one input and one output layer, and a varying number of perceptrons in each layer. 

	Hence, the QNN is a quantum circuit of quantum perceptrons that acts on an initial state $\rho^\mathrm{in}$ of the input qubits and puts out an, in general, mixed state $\rho^\mathrm{out}$ for the output qubits. Consider a network with $L$ hidden layers as well as an input and an output layer. The output state of the network is then
		\begin{equation}
		\label{eq:rhoout}
			\rho^\mathrm{out}=\mathrm{tr}_\mathrm{in,hidden}\left(U^\mathrm{out}U^L\dots U^1\left(\rho^\mathrm{in}\otimes\lvert 0\dots 0\rangle_\mathrm{hidden,out}\langle 0\dots 0\rvert\right){U^1}^\dagger\dots{U^L}^\dagger {U^\mathrm{out}}^\dagger\right),
		\end{equation}
	where $U^l$ are the layer unitaries, which are comprised of a product of quantum perceptrons acting on the qubits in layer $l-1$ and $l$:
		\begin{equation*}
			U^l=U_{m_l}^lU_{m_l-1}^l\dots U_1^l,
		\end{equation*}	
	where $m_l$ is the number of qubits in layer $l$. Note that since we allow arbitrary unitary operators, the perceptrons do not, in general, commute. Due to the structure of the proposed QNN, the network output can be expressed as the composition of a sequence of completely positive layer-to-layer transition maps $\mathcal{E}^l$:
		\begin{align*}
			\rho_x^\mathrm{out}(s)&=\mathcal{E}_s^{\mathrm{out}}\left(\mathcal{E}_s^{L}\left(\dots\mathcal{E}_s^{2}\left(\mathcal{E}_s^{1}\left(\rho_x^\mathrm{in}\right)\right)\dots\right)\right)
		\end{align*}
	with the channel going from layer $l-1$ to $l$ being
		\begin{align}
		\label{eq:channel}
			\mathcal{E}_s^{l}\left(X^{l-1}\right)&=\mathrm{tr}_{l-1}\left(U_{m_l}^l(s)\dots U_1^l(s)\left(X^{l-1}\otimes\lvert 0\dots 0\rangle_l\langle 0\dots 0\rvert\right){U_1^l}^\dagger(s)\dots{U_{m_l}^l}^\dagger(s)\right),
		\end{align}
	where $m_l$ is the number of perceptrons in layer $l$. This implies that the action of the network on the input state can be computed layer by layer, such that we never have to store the state of the whole network.  Here we let the perceptron unitaries to depend on some sort of time parameter $s$. Training the network then corresponds to finding a path of unitaries $U(s)$ that eventually minimise the cost function. We achieve this using the following update rule for the unitary after a time step $\epsilon$:
		\begin{equation}
		\label{eq:updateU}
			U_j^l(s+\epsilon)=e^{i\epsilon K_j^l(s)} U_j^l(s),
		\end{equation}
	where $K_j^l(s)$ is the matrix that includes all parameters of the $j^\mathrm{th}$ perceptron unitary in layer $l$.  We will explain how to compute $K_j^l(s)$ in sections \ref{sec:TrainingQNN} and \ref{sec:QuantumNNopt}.

\section{Universality and implementing quantum channels}
\label{sec:universality}

It is known (see e.g.\ \cite{Wolf2018}) that a neural network composed of classical perceptrons can represent any function. Hence, it is desirable to have the same feature for quantum neural networks.

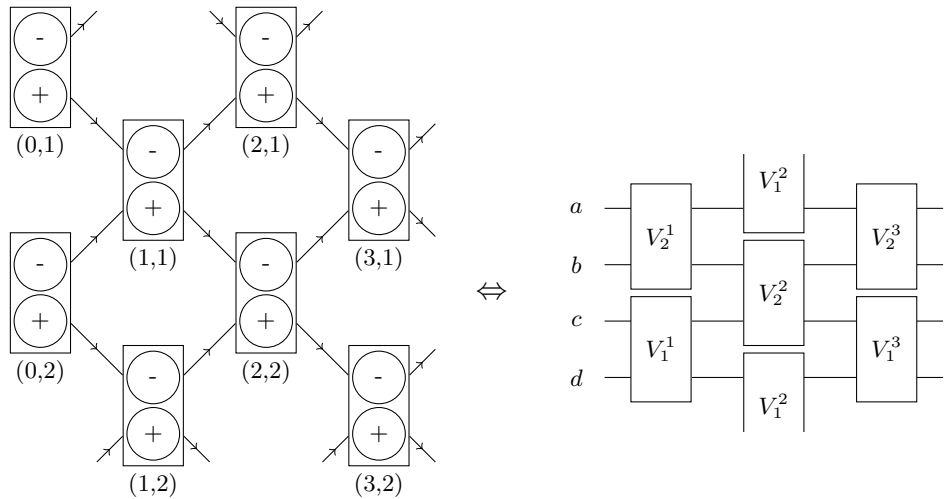
\begin{figure}[H]
	\centering
\begin{tikzpicture}[scale=1.5,decoration={markings,mark=at position 0.5 with {\arrow{>}}} ]
\node[draw,minimum width=.8cm,minimum height=1.6cm] at (0,1) (a) {};
\node[draw,minimum width=.8cm,minimum height=1.6cm]  at (0,3) (b) {};
\node[draw,minimum width=.8cm,minimum height=1.6cm]  at (1,0) (c) {};
\node[draw,minimum width=.8cm,minimum height=1.6cm]  at (1,2) (d) {};
\node[draw,minimum width=.8cm,minimum height=1.6cm]  at (2,1) (e) {};
\node[draw,minimum width=.8cm,minimum height=1.6cm]  at (2,3) (f) {};
\node[draw,minimum width=.8cm,minimum height=1.6cm]  at (3,0) (g) {};
\node[draw,minimum width=.8cm,minimum height=1.6cm]  at (3,2) (h) {};
\node[draw,circle,minimum size=.7cm] at ([yshift=.25cm]a){-};
\node[draw,circle,minimum size=.7cm] at ([yshift=-.25cm]a){+};
\node at ([yshift=-.7cm]a){(0,2)};
\node[draw,circle,minimum size=.7cm] at ([yshift=.25cm]b){-};
\node[draw,circle,minimum size=.7cm] at ([yshift=-.25cm]b){+};
\node at ([yshift=-.7cm]b){(0,1)};
\node[draw,circle,minimum size=.7cm] at ([yshift=.25cm]c){-};
\node[draw,circle,minimum size=.7cm] at ([yshift=-.25cm]c){+};
\node at ([yshift=-.7cm]c){(1,2)};
\node[draw,circle,minimum size=.7cm] at ([yshift=.25cm]d){-};
\node[draw,circle,minimum size=.7cm] at ([yshift=-.25cm]d){+};
\node at ([yshift=-.7cm]d){(1,1)};
\node[draw,circle,minimum size=.7cm] at ([yshift=.25cm]e){-};
\node[draw,circle,minimum size=.7cm] at ([yshift=-.25cm]e){+};
\node at ([yshift=-.7cm]e){(2,2)};
\node[draw,circle,minimum size=.7cm] at ([yshift=.25cm]f){-};
\node[draw,circle,minimum size=.7cm] at ([yshift=-.25cm]f){+};
\node at ([yshift=-.7cm]f){(2,1)};
\node[draw,circle,minimum size=.7cm] at ([yshift=.25cm]g){-};
\node[draw,circle,minimum size=.7cm] at ([yshift=-.25cm]g){+};
\node at ([yshift=-.7cm]g){(3,2)};
\node[draw,circle,minimum size=.7cm] at ([yshift=.25cm]h){-};
\node[draw,circle,minimum size=.7cm] at ([yshift=-.25cm]h){+};
\node at ([yshift=-.7cm]h){(3,1)};
\draw[postaction={decorate}] (a) -- (c);
\draw[postaction={decorate}] (c) -- (e);
\draw[postaction={decorate}] (e) -- (g);
\draw[postaction={decorate}] (a) -- (d);
\draw[postaction={decorate}] (d) -- (e);
\draw[postaction={decorate}] (e) -- (h);
\draw[postaction={decorate}] (b) -- (d);
\draw[postaction={decorate}] (d) -- (f);
\draw[postaction={decorate}] (f) -- (h);
\draw[postaction={decorate}] (.5,-.5) -- (c);
\draw[postaction={decorate}] (c) -- (1.5,-.5);
\draw[postaction={decorate}] (2.5,-.5)-- (g);
\draw[postaction={decorate}] (b) -- (.5,3.5);
\draw[postaction={decorate}] (1.5,3.5) -- (f);
\draw[postaction={decorate}] (f)--(2.5,3.5);
\draw[postaction={decorate}] (g)--(3.5,-.5);
\draw[postaction={decorate}] (g)--(3.5,.5);
\draw[postaction={decorate}] (h)--(3.5,1.5);
\draw[postaction={decorate}] (h)--(3.5,2.5);
\node at (4,1) (v) {\large$\Leftrightarrow$};
\begin{scope}[shift={(5.5,.5)}]
\draw (-.5,-.25)--(2.5,-.25);
\draw (-.5,.25)--(2.5,.25);
\draw (-.5,.75)--(2.5,.75);
\draw (-.5,1.25)--(2.5,1.25);
\node[draw,fill=white,minimum width=.8cm,minimum height=1.4cm] at (0,0) (a) {$V_1^1$};
\node[draw,fill=white,minimum width=.8cm,minimum height=1.4cm]  at (0,1) (b) {$V_2^1$};
\node[draw,fill=white,minimum width=.8cm,minimum height=1.4cm]  at (1,.5) (c) {$V_2^2$};
\node[draw,fill=white,minimum width=.8cm,minimum height=1.4cm]  at (1,1.5) (d) {$V_1^2$};
\node[draw,fill=white,minimum width=.8cm,minimum height=1.4cm]  at (2,0) (e) {$V_1^3$};
\node[draw,fill=white,minimum width=.8cm,minimum height=1.4cm]  at (2,1) (f) {$V_2^3$};
\node[draw,fill=white,minimum width=.8cm,minimum height=1.4cm]  at (1,-.5) (g) {$V_1^2$};
\node[fill=white,minimum width=1cm,minimum height=.8cm]  at (1,-1) (g) {};
\node[fill=white,minimum width=1cm,minimum height=.8cm]  at (1,2) (g) {};
\node[] at (-.75,-.25) (a) {$d$};
\node[]  at (-.75,.25) (b) {$c$};
\node[]  at (-.75,.75) (c) {$b$};
\node[]  at (-.75,1.25) (d) {$a$};
\end{scope}
\end{tikzpicture}
	\caption{Universality of the quantum neural network.  The $+$ qubit in neuron $(0,1)$ on the left hand side diagram corresponds to the qubit labelled by $a$ on the right hand side.  Similarly, the $(0,2)$ $-$ qubit corresponds to the qubit labelled by $b$ and so on.}
	\label{fig:universality}
\end{figure}

In order to show universality, let us construct a particular network that is capable of universal quantum computation. For this we number neurons by two indices: neuron $(l,j)$ is the $j$th neuron in $l$th layer. Let there be $m_l$ neurons in $l$th layer. Consider a network where the neuron $(l,j)$ is connected to neurons $(l-1,j)$ and $\left(l+1,j+(-1)^l\mod m_l,\right)$ for all $(l,j)$ and no other connections exist. Suppose that each neuron corresponds to two qubits, labelled by $+$ and $-$, initialised as $|00\rangle$ (as shown in the left picture of Figure~\ref{fig:universality}). The action of the neural network on one layer has the form
	\begin{equation}\label{eq:qcdecomp}
		\rho^l=\mathrm{tr}_{l-1}\left(U^l\left(\rho^{l-1}\otimes\left[\bigotimes_{j=1}^{m_l}\lvert 00\rangle_{(l,j)}\langle 00\rvert\right]\right){U^l}^\dagger\right),
	\end{equation}
	where $U^l=\prod_{j=m_l}^{1}U^l_j$ is a product of each unitary perceptron acting on layers $l$ and $l+1$.
For the neuron $(l,j)$, choose 
	\begin{equation*}
		U^l_j = V_j^l\ \mathrm{SWAP}\left[\left(l-1,j,-\right),\left(l,j-\frac{1+(-1)^l}{2},+\right)\right] \mathrm{SWAP}\left[\left(l-1,j,+\right),\left(l,j+\frac{1-(-1)^l}{2},-\right)\right],
	\end{equation*}
where the SWAP operators act on one qubit in the $l-1$th layer and one qubit that correspond to the neuron $(l,j)$ and $V_j^l$ is a unitary that acts on the qubits of the neuron $(l,j)$.  For example, the first swap swaps the $-$ qubit in the neuron $(l,j)$ with the $+$ qubit in neuron $(l,j-(1+(-1)^l)/2)$.  Note that for fixed $l$ all swaps commute since they all act on different pairs of qubits.  This neural network is equivalent to the quantum circuit of two-qubit gates $V_j^l$ that act on registers number $2j-1$ and $2j$ at the $l$th time step. This quantum circuit is universal, as two-qubit gates are universal (see e.g.\ \cite{Nielsen2010}) and SWAP is one of them (see Figure~\ref{fig:universality}).  (Note that one could easily consider different geometries for the network to allow far away qubits to interact, which may be useful for simulating certain quantum circuits more efficiently.)

It is also straightforward to see that the most general form of a quantum perceptron we allow can implement any quantum channel on the input qubits (or qudits if we are dealing with more general neurons).  To see this, look at equation (\ref{eq:qcdecomp}), and suppose that $2m_{l-1}=m_l$.  Then it follows from the Stinespring dilation theorem \cite{Werner2001} that, because the layer $l$ qubits are in a pure state, we can choose $U^l$ to implement any completely positive map we like on the $l-1$ qubits.  Note that the output state lives on the $l$ system as opposed to the $l-1$ systems.  This is equivalent to the usual Stinespring protocol by choosing $S U^l=\tilde{U}^l$, where $\tilde{U}^l$ implements the channel we want on the $l-1$ qubits and $S$ swaps these qubits into the first $m_{l-1}$ qubits of the $l$ layer.  Of course, this is just a proof of principle.  In realistic cases, we would not want to consider generic unitaries $U^l$ that act on $m_{l-1}+m_l$ qubits, but rather we want to choose $U^l=\prod_{j=1}^{m_l}U^l_j$, where each $U^l_j$ acts only on a few qubits.  This would be much easier to implement in practice.  Then it is an interesting question which channels can be simulated by these more restricted class of perceptrons.

\section{Classical simulation of training the QNN}
\label{sec:TrainingQNN}

In this section, we describe how the simulation of the proposed QNN can be done on a classical computer.
\subsection{Example: A Simple Network}

	\label{subsec:simplenetwork}	
	\begin{figure}[H]
	\centering
	\includegraphics[width=.25\linewidth]{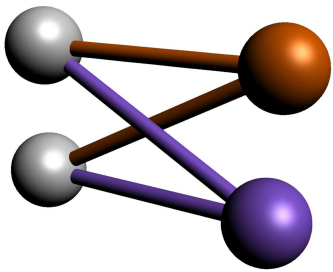}
	\caption{Simple QNN with two layers (i.e.\ no hidden layers). We apply the unitaries from bottom to top: first the orange one and the violet one afterwards.}
	\label{fig:22QNN}
	\end{figure}
To clarify how the training of a QNN works, we consider a simple example of a two-layer network with four qubits in total, as shown in Figure~\ref{fig:22QNN}.

	The algorithm is as follows:
		\begin{itemize}
			\item[I.] Initialise:
			\begin{itemize}
				\item[I.1] Set $s=0$.
				\item[I.2] Choose $U_1^\mathrm{out}(0)$ and $U_2^\mathrm{out}(0)$ at random.
			\end{itemize}
			\item[II.]  Feedforward: For each element $\left(\lvert\phi^\mathrm{in}_x\rangle,\lvert\phi^\mathrm{out}_x\rangle\right)$ in the set of training data, do the following steps:
			\begin{itemize}
				\item[II.1] Initialise the network in the state
				\begin{equation*}
						\lvert\Phi^\mathrm{in}_x\rangle=\lvert \phi^\mathrm{in}_x\rangle\otimes\lvert 00\rangle_\mathrm{out}.
				\end{equation*}
				\item[II.2] Apply the unitaries to the input state:
					\begin{equation*}
						\lvert\Psi_x\rangle=U_2^\mathrm{out}(s)U_1^\mathrm{out}(s)\lvert\Phi_x\rangle.
					\end{equation*}
				\item[II.3] Trace out the input system:
					\begin{equation*}
						\rho_x^\mathrm{out}(s)=\mathrm{tr}_\mathrm{in}\left(\lvert\Psi_x\rangle\langle\Psi_x\rvert\right).
					\end{equation*}
			\end{itemize}
			\item [III.] Update the parameters:
			\begin{itemize}
				\item[III.1] Compute the cost function:
					\begin{equation*}
						C(s)=\frac{1}{N} \sum_{x=1}^N \langle\psi_x\rvert\rho_x^\mathrm{out}(s)\lvert\psi_x\rangle.		
					\end{equation*}
				\item[III.2] Calculate the parameter matrices $K_j^l(s)$.  (How to do this is explained below.)
				\item[III.3] Update each perceptron unitary via 
					\begin{equation*}
						U_j^l(s+\epsilon)=e^{i\epsilon K_j^l(s)} U_j^l(s).
					\end{equation*}
				\item[III.4] Update $s=s+\epsilon$.
			\end{itemize}
			\item[IV.] Repeat steps II. and III. until the cost function has reached its maximum.
		\end{itemize}
		
	To perform the algorithm, we need a formula that allows us to compute the parameter matrices $K_j^l(s)$ to update the perceptron unitaries, which we will derive in the following. For clarity, we omit the superscript that indicates the layer since there is only one layer of unitaries. Furthermore, for the unitaries we omit the dependence on $s$ for reasons of clarity. We derive the formula for the parameter matrices $K_j^l(s)$ as follows: Consider the derivative of the cost function,
		\begin{equation}
			\frac{dC}{ds}=\lim_{\epsilon\rightarrow 0}\frac{C(s+\epsilon)-C(s)}{\epsilon}.
			\label{eq:costderivative}
		\end{equation}
	In the following calculation, we have the convention that all unitaries act on the whole system, e.g.\ $U_1$ is actually $U_1\otimes\mathbb{I}_{\mathrm{qubit}_2^2}$. To calculate Equation \ref{eq:costderivative}, we need the output state for the updated unitary $U(s+\epsilon)=e^{i\epsilon K_2}U_2\ e^{i\epsilon K_1}U_1$, which is
		\begin{align*}
			\rho_x^\mathrm{out}(s+\epsilon)&=\mathrm{tr}_\mathrm{in}\left(e^{i\epsilon K_2}U_2\ e^{i\epsilon K_1}U_1\ \lvert\Phi^\mathrm{in}_x\rangle\langle\Phi^\mathrm{in}_x\rvert\ U_1^\dagger e^{-i\epsilon K_1}\ U_2^\dagger e^{-i\epsilon K_2}\right)\\
			&=\rho_x^\mathrm{out}(s)+i\epsilon\ \mathrm{tr}_\mathrm{in}\left(U_2  K_1 U_1 \ \lvert\Phi_x\rangle\langle\Phi^\mathrm{in}_x\rvert\ U_1^\dagger U_2^\dagger -U_2  U_1 \ \lvert\Phi_x\rangle\langle\Phi^\mathrm{in}_x\rvert\ U_1^\dagger  K_1 U_2^\dagger \right.\\
			&\hspace{15pt}\left.+K_2 U_2  U_1 \ \lvert\Phi_x\rangle\langle\Phi_x\rvert\ U_1^\dagger U_2^\dagger -U_2  U_1 \ \lvert\Phi^\mathrm{in}_x\rangle\langle\Phi^\mathrm{in}_x\rvert\ U_1^\dagger U_2^\dagger  K_2 \right)+\mathcal{O}(\epsilon^2).
		\end{align*}
	Hence, the derivation of the cost function becomes
		\begin{align*}
			\frac{dC}{ds}&=\lim_{\epsilon\rightarrow 0}\frac{C(s)+i\epsilon\frac{1}{N} \sum_x\langle\phi^\mathrm{out}_x\rvert\mathrm{tr}_\mathrm{in}(\dots)\lvert\phi^\mathrm{out}_x\rangle-C(s)}{\epsilon}\\
			&=\frac{i}{N}\ \sum_{x=1}^N\mathrm{tr}\left(\mathbb{I}_\mathrm{in}\otimes \lvert\phi^\mathrm{out}_x\rangle\langle\phi^\mathrm{out}_x\rvert\left(U_2\left[K_1,U_1\lvert\Phi^\mathrm{in}_x\rangle\langle\Phi^\mathrm{in}_x\rvert U_1^\dagger\right]U_2^\dagger+\left[K_2,U_2U_1\lvert\Phi^\mathrm{in}_x\rangle\langle\Phi^\mathrm{in}_x\rvert U_1^\dagger U_2^\dagger\right]\right) \right)\\
			&=\frac{i}{N}\ \sum_{x=1}^N\mathrm{tr}\Big(\underbrace{\left[U_1\lvert\Phi^\mathrm{in}_x\rangle\langle\Phi^\mathrm{in}_x\rvert U_1^\dagger,U_2^\dagger\left(\mathbb{I}_\mathrm{in}\otimes \lvert\phi^\mathrm{out}_x\rangle\langle\phi^\mathrm{out}_x\rvert\right)U_2\right]}_{\equiv M_1}K_1+\underbrace{\left[U_2U_1\lvert\Phi^\mathrm{in}_x\rangle\langle\Phi^\mathrm{in}_x\rvert U_1^\dagger U_2^\dagger,\mathbb{I}_\mathrm{in}\otimes \lvert\psi_x\rangle\langle\psi_x\rvert\right]}_{\equiv M_2}K_2\Big)\\
			&=\frac{i}{N}\ \sum_{x=1}^N\mathrm{tr}\left(M_1K_1+M_2K_2\right).
		\end{align*}
	We will parametrise the parameter matrices as
		\begin{equation*}
			K_1(s)=\sum_{\alpha\beta\gamma}K_{1,\alpha\beta\gamma}(s)\left(\sigma^\alpha\otimes\sigma^\beta\otimes\sigma^\gamma\right),
		\end{equation*}
	where $\sigma\equiv\{\mathbb{I},\sigma^x,\sigma^y,\sigma^z\}$, since every unitary $U_j^l$ in this example acts on three qubits.
	To reach the maximum of the cost function as a function of the parameters \emph{fastest}, we maximize $\frac{dC}{ds}$. Since this is a linear function (up to order $\epsilon$), the extrema are at $\pm\infty$. To ensure that we get a finite solution we introduce a Lagrange multiplier $\lambda\in\mathbb{R}$. Hence, to find $K_1$ we have to solve the following maximization problem:
		\begin{align*}
			\max_{K_{1,\alpha\beta\gamma}}\left(\frac{dC}{ds}-\lambda\sum_{\alpha\beta\gamma}{K_{1,\alpha\beta\gamma}}^2\right)&=\max_{K_{1,\alpha\beta\gamma}}\left(\frac{i}{N} \sum_x\mathrm{tr}\left(M_1 \sum_{\alpha\beta\gamma}K_{1,\alpha\beta\gamma}\left(\sigma^\alpha\otimes\sigma^\beta\otimes\sigma^\gamma\right)\otimes\mathbb{I}_{\mathrm{qubit}_2^2}+M_2K_2\right)-\lambda \sum_{\alpha\beta\gamma}{K_{1,\alpha\beta\gamma}}^2\right)\\
																		 &=\max_{K_{1,\alpha\beta\gamma}}\left(\frac{i}{N} \sum_x\mathrm{tr}_{1,2,3}\left(\mathrm{tr}_4 \left(M_1\right) \sum_{\alpha\beta\gamma}K_{1,\alpha\beta\gamma}\left(\sigma^\alpha\otimes\sigma^\beta\otimes\sigma^\gamma\right)+\mathrm{tr}_4\left(M_2K_2\right)\right)\right.\\&\hspace{15pt}\left.-\lambda \sum_{\alpha\beta\gamma}{K_{1,\alpha\beta\gamma}}^2\right).
		\end{align*}
	Taking the derivative with respect to $K_{1,\alpha\beta\gamma}$ yields
		\begin{equation*}
			\frac{i}{N}\sum_x\mathrm{tr}_{1,2,3}\left(\mathrm{tr}_4\left(M_1\right)\left(\sigma^\alpha\otimes\sigma^\beta\otimes\sigma^\gamma\right)\right)-2\lambda K_{1,\alpha\beta\gamma}=0.
		\end{equation*}
	Therefore, the elements of the parameter matrix are
		\begin{equation*}
			K_{1,\alpha\beta\gamma}=\frac{i}{2\lambda N}\  \sum_x\mathrm{tr}_{1,2,3}\left(\mathrm{tr}_4\left(M_1\right)\left(\sigma^\alpha\otimes\sigma^\beta\otimes\sigma^\gamma\right)\right).
		\end{equation*}
	This yields the matrix 
		\begin{align*}
			K_1&=\sum_{\alpha\beta\gamma}K_{1,\alpha\beta\gamma}\left(\sigma^\alpha\otimes\sigma^\beta\otimes\sigma^\gamma\right)\\
			&=\frac{i}{2\lambda N}\sum_x\sum_{\alpha\beta\gamma}\mathrm{tr}_{1,2,3}\left(\mathrm{tr}_4\left(M_1\right)\left(\sigma^\alpha\otimes\sigma^\beta\otimes\sigma^\gamma\right)\right)\left(\sigma^\alpha\otimes\sigma^\beta\otimes\sigma^\gamma\right)\\
			&=\frac{4 i}{\lambda N}\sum_x\mathrm{tr}_4\left(M_1\right).
		\end{align*}
	In the last step we have used the completeness relation for the Pauli matrices.	This derivation works analogously for $K_2$, which is
		\begin{equation*}
			K_2=\frac{4 i}{\lambda N}\sum_x\mathrm{tr}_3\left(M_2\right).
		\end{equation*}
	Using these formulas for step III.3 of the algorithm, we can train the QNN. Note that in the paper, we have introduced the learning rate $\eta$, which is related to lambda by $\eta=1/\lambda$ and referred to it as the \emph{learning rate}. For all numerical computations we show here, we will always indicate which $\eta$ we have used to make it comparable to the plots in the main paper.
	
	In Figure~\ref{fig:cost2-2}, the cost function is depicted for different values of the parameter $\eta$. To generate this figure, we have used a training set of $10$ randomly generated pairs and $\epsilon=0.1$.
\input{FigureCost}	
	
\subsection{The General Network}
	
	We will now generalise the previous example to the training of arbitrary networks. The training algorithm then is as follows:

	\begin{itemize}
		\item[I.] Initialise:
			\begin{itemize}
				\item[I.1] Set $s=0$.
				\item[I.2] Choose all $U_j^l(0)$ randomly.
			\end{itemize}
		\item[II.] Feedforward: For each element $\left(\lvert\phi^\mathrm{in}_x\rangle,\lvert\phi^\mathrm{out}_x\rangle\right)$ in the set of training data, do the following steps: For every layer $l$, do the following:
			\begin{itemize}
				\item[II.1] Tensor the state of the layer to the output state of layer $l-1$, where $\rho_x^\mathrm{in}=\lvert\phi^\mathrm{in}_x\rangle\langle\phi^\mathrm{in}_x\rvert$:
					\begin{equation*}
						\rho_x^{l-1}(s)\otimes\lvert 0\dots 0\rangle_l\langle 0\dots 0\rvert
					\end{equation*}
				\item[II.2] Apply the unitaries in layer $l$:
					\begin{equation*}
						U_{m(l)}^l(s) U_{m(l)-1}^l(s)\dots U_1^l(s)\left(\rho_x^{l-1}(s)\otimes\lvert 0\dots 0\rangle_l\langle 0\dots 0\rvert\right) {U_1^l}^\dagger(s)\dots{U_{m(l)-1}^l}^\dagger(s) {U_{m(l)}^l}^\dagger(s)
					\end{equation*}
				\item[II.3] Trace out layer $l-1$:
					\begin{equation*}
						\rho_x^l(s)=\mathrm{tr}_{l-1}\left(U_{m(l)}^l(s) U_{m(l)-1}^l(s)\dots U_1^l(s)\left(\rho_x^{l-1}(s)\otimes\lvert 0\dots 0\rangle_l\langle 0\dots 0\rvert\right) {U_1^l}^\dagger(s)\dots{U_{m(l)-1}^l}^\dagger(s) {U_{m(l)}^l}^\dagger(s)\right).
					\end{equation*}
				\item[II.4] Store $\rho_x^l(s)$. This step is crucial to efficiently calculate the parameter matrices.
			\end{itemize}
		These steps are equivalent to applying the layer-to-layer channels $\mathcal{E}_s^{l}$ defined in \cref{eq:channel} successively to the input state. 
				
		\item[III.] Update parameters:
			\begin{itemize}
				\item[III.1] Compute the cost function: 
					\begin{equation*}
						C(s)=\frac{1}{N}\sum_{x=1}^N\langle\phi^\mathrm{out}_x\rvert\rho_x^\mathrm{out}(s)\lvert\phi^\mathrm{out}_x\rangle
					\end{equation*}
				\item[III.2] Calculate each parameter matrix $K_j^l(s)$.  (How to do this is explained below.)
				\item[III.3] Update each perceptron unitary via
					\begin{equation*}
						U_j^l(s+\epsilon)=e^{i\epsilon K_j^l(s)}U_j^l(s).
					\end{equation*}
				\item[III.4] Update $s=s+\epsilon$.
			\end{itemize}
		\item[IV.] Repeat steps II. and III. until the cost function has reached its maximum.
	\end{itemize}

	We will now generalise the derivation of the update matrices $K_j^l(s)$ given in \cref{subsec:simplenetwork}. As above, the unitaries always act on the current layers, e.g.\ $U^2_1$ is actually $U^2_1\otimes \mathbb{I}^2_{2,3,\dots m_2}$. Let $\rho_x^\mathrm{in}=\lvert\phi^\mathrm{in}_x\rangle\langle\phi^\mathrm{in}_x\rvert$. The output state at step $s+\epsilon$ is then
		\begin{align*}
			\rho_x^\mathrm{out}(s+\epsilon)&=\mathrm{tr}_\mathrm{in,hidden}\left(e^{i\epsilon K_{m_\mathrm{out}}^\mathrm{out}(s)}U_{m_\mathrm{out}}^\mathrm{out}(s)\ e^{i\epsilon K_{m_\mathrm{out}-1}^\mathrm{out}(s)}U_{m_\mathrm{out}-1}^\mathrm{out}(s)\dots e^{i\epsilon K_1^1(s)}U_1^1(s)\left(\rho_x^\mathrm{in}\otimes\lvert 0\dots 0\rangle_\mathrm{hidden,out}\langle 0\dots 0\rvert\right)\right.\\
			&\hspace{15pt}\left.{U_1^1}^\dagger(s)e^{-i\epsilon K_{1}^1(s)}\dots{U_{m_\mathrm{out}-1}^\mathrm{out}}^\dagger(s)e^{-i\epsilon K_{m_\mathrm{out}-1}^\mathrm{out}(s)} \ {U_{m_\mathrm{out}}^\mathrm{out}}^\dagger(s)e^{-i\epsilon K_{m_\mathrm{out}}^\mathrm{out}(s)}\right)\\
			&=\rho_x^\mathrm{out}(s)+i\epsilon\ \mathrm{tr}_\mathrm{in,hidden}\left(K_{m_\mathrm{out}}^\mathrm{out} U_{m_\mathrm{out}}^\mathrm{out}\dots U_1^1(s)\left(\rho_x^\mathrm{in}\otimes\lvert 0\dots 0\rangle_\mathrm{hidden,out}\langle 0\dots 0\rvert\right){U_1^1}^\dagger(s)\dots{U_{m_\mathrm{out}}^\mathrm{out}}^\dagger(s)\right.\\
			&\hspace{15pt}\left.-U_{m_\mathrm{out}}^\mathrm{out}\dots U_1^1(s)\left(\rho_x^\mathrm{in}\otimes\lvert 0\dots 0\rangle_\mathrm{hidden,out}\langle 0\dots 0\rvert\right){U_1^1}^\dagger(s)\dots{U_{m_\mathrm{out}}^\mathrm{out}}^\dagger(s)K_{m_\mathrm{out}}^\mathrm{out}(s)+\dots\right.\\
			&\hspace{15pt}\left.+U_{m_\mathrm{out}}^\mathrm{out}\dots K_1^1(s)U_1^1(s)\left(\rho_x^\mathrm{in}\otimes\lvert 0\dots 0\rangle_\mathrm{hidden,out}\langle 0\dots 0\rvert\right){U_1^1}^\dagger(s)\dots{U_{m_\mathrm{out}}^\mathrm{out}}^\dagger(s)\right.\\
			&\hspace{15pt}\left.-U_{m_\mathrm{out}}^\mathrm{out}\dots U_1^1(s)\left(\rho_x^\mathrm{in}\otimes\lvert 0\dots 0\rangle_\mathrm{hidden,out}\langle 0\dots 0\rvert\right){U_1^1}^\dagger(s)K_1^1(s)\dots{U_{m_\mathrm{out}}^\mathrm{out}}^\dagger(s)\right)+\mathcal{O}\left(\epsilon^2\right)\\
			&=\rho_x^\mathrm{out}(s)+i\epsilon\  \mathrm{tr}_\mathrm{in,hidden}\left(\left[K_{m_\mathrm{out}}^\mathrm{out}(s), U_{m_\mathrm{out}}^\mathrm{out}(s)\dots U_1^1(s)\left(\rho_x^\mathrm{in}\otimes\lvert 0\dots 0\rangle_\mathrm{hidden,out}\langle 0\dots 0\rvert\right){U_1^1}^\dagger(s)\dots{U_{m_\mathrm{out}}^\mathrm{out}}^\dagger(s)\right]+\dots\right.\\
			&\hspace{15pt}\left.+U_{m_\mathrm{out}}^\mathrm{out}(s)\dots U_2^1(s) \left[K_1^1(s),U_1^1(s)\left(\rho_x^\mathrm{in}\otimes\lvert 0\dots 0\rangle_\mathrm{hidden,out}\langle 0\dots 0\rvert\right){U_1^1}^\dagger(s)\right]{U_2^1}^\dagger(s)\dots{U_{m_\mathrm{out}}^\mathrm{out}}^\dagger(s)\right)+\mathcal{O}\left(\epsilon^2\right).
 		\end{align*}
	The derivative of the cost function up to first order in $\epsilon$ can then be written as
		\begin{align}
		\label{eq:taylordC}
		\begin{split}
				\frac{dC(s)}{ds}&=\lim_{\epsilon\rightarrow 0}\frac{C(s+\epsilon)-C(s)}{\epsilon}\\
				&=\lim_{\epsilon\rightarrow 0}\frac{C(s)+\frac{i\epsilon}{N} \sum_x\langle\phi^\mathrm{out}_x\rvert\mathrm{tr}_\mathrm{in,hidden}\left(\rho_x^\mathrm{out}(s+\epsilon)\right)\lvert\phi^\mathrm{out}_x\rangle-C(s)}{\epsilon}\\
				&=\frac{1}{N}\sum_x \mathrm{tr}\left(\mathbb{I}_\mathrm{in,hidden}\otimes\lvert\phi^\mathrm{out}_x\rangle\langle\phi^\mathrm{out}_x\rvert\left(\left[iK_{m_\mathrm{out}}^\mathrm{out}(s), U_{m_\mathrm{out}}^\mathrm{out}(s)\dots U_1^1(s)\left(\rho_x^\mathrm{in}\otimes\lvert 0\dots 0\rangle_\mathrm{hidden,out}\langle 0\dots 0\rvert\right){U_1^1}^\dagger(s)\right.\right.\right.\\
				&\hspace{15pt}\left.\left.\left.\dots{U_{m_\mathrm{out}}^\mathrm{out}}^\dagger(s)\right]+\dots+U_{m_\mathrm{out}}^\mathrm{out}(s)\dots U_2^1(s) \left[iK_1^1(s),U_1^1(s)\left(\rho_x^\mathrm{in}\otimes\lvert 0\dots 0\rangle_\mathrm{hidden,out}\langle 0\dots 0\rvert\right){U_1^1}^\dagger(s)\right]\right.\right.\\
				&\hspace{15pt}\left.\left.{U_2^1}^\dagger(s)\dots{U_{m_\mathrm{out}}^\mathrm{out}}^\dagger(s)\right)\right)\\
				&=\frac{1}{N}\sum_x \mathrm{tr}\Big(\underbrace{\left[U_{m_\mathrm{out}}^\mathrm{out}(s)\dots\left(\rho_x^\mathrm{in}\otimes\lvert 0\dots 0\rangle_\mathrm{hidden,out}\langle 0\dots 0\rvert\right)\dots{U_{m_\mathrm{out}}^\mathrm{out}}^\dagger(s),\mathbb{I}_\mathrm{in,hidden}\otimes\lvert\phi^\mathrm{out}_x\rangle\langle\phi^\mathrm{out}_x\rvert\right]}_{\equiv M_{m_\mathrm{out}}^\mathrm{out}(s)}iK_{m_\mathrm{out}}^\mathrm{out}(s)+\dots\\
				&\hspace{15pt}+\underbrace{\left[U_1^1(s)\left(\rho_x^\mathrm{in}\otimes\lvert 0\dots 0\rangle_\mathrm{hidden,out}\langle 0\dots 0\rvert\right){U_1^1}^\dagger(s),{U_2^1}^\dagger(s)\dots{U_{m_\mathrm{out}}^\mathrm{out}}^\dagger(s)\left(\mathbb{I}_\mathrm{in+hidden}\otimes\lvert\psi_x\rangle\langle\psi_x\rvert\right)U_{m_\mathrm{out}}^\mathrm{out}(s)\dots U_2^1(s)\right]}_{\equiv M_1^1(s)}\\&\hspace{15pt}iK_1^1(s)\Big)\\
				&=\frac{i}{N}\sum_x \mathrm{tr}\left(M_{m_\mathrm{out}}^\mathrm{out}(s) K_{m_\mathrm{out}}^\mathrm{out}(s)+\ \dots\ +M_1^1(s)K_1^1(s)\right).
		\end{split}
		\end{align}
	We will parametrise the parameter matrices as
		\begin{equation*}
			K_j^l(s)=\sum_{\alpha_1,\alpha_2,\dots,\alpha_{m_{l-1}},\beta}K^l_{j,\alpha_1,\dots,\alpha_{m_{l-1}},\beta}(s)\left(\sigma^{\alpha_1}\otimes\ \dots\ \otimes\sigma^{\alpha_{m_{l-1}}}\otimes\sigma^\beta\right),
		\end{equation*}
	where the $\alpha_i$ denote the qubits in the previous layer and $\beta$ denotes the current qubit in layer $l$. As described in the example, to reach the maximum of the cost function as a function of the parameters \emph{fastest}, we maximize $\frac{dC}{ds}$. Since this is a linear function, the extrema are at $\pm\infty$. To ensure that we get a finite solution, we introduce a Lagrange multiplier $\lambda\in\mathbb{R}$. Hence, to find $K_j^l$ we have to solve the following maximization problem:
		\begin{align*}
			\max_{K^l_{j,\alpha_1,\dots,\beta}}&\left(\frac{dC(s)}{ds}-\lambda\sum_{\alpha_i,\beta}{K^l_{j,\alpha_1,\dots,\beta}}(s)^2\right)\\
			&=\max_{K^l_{j,\alpha_1,\dots,\beta}}\left(\frac{i}{N}\sum_x \mathrm{tr}\left(M_{m_\mathrm{out}}^\mathrm{out}(s) K_{m_\mathrm{out}}^\mathrm{out}(s)+\ \dots\ +M_1^1(s)K_1^1(s)\right)-\lambda\sum_{\alpha_1,\dots,\beta}{K^l_{j,\alpha_1,\dots,\beta}}(s)^2\right)\\
			&=\max_{K^l_{j,\alpha_1,\dots,\beta}}\left(\frac{i}{N}\sum_x\mathrm{tr}_{\alpha_1,\dots,\beta}\left(\mathrm{tr}_\mathrm{rest}\left(M_{m_\mathrm{out}}^\mathrm{out}(s) K_{m_\mathrm{out}}^\mathrm{out}(s)+\ \dots\ +M_1^1(s)K_1^1(s)\right)\right)-\lambda\sum_{\alpha_1,\dots,\beta}{K^l_{j,\alpha_1,\dots,\beta}}(s)^2\right).
		\end{align*}
	Taking the derivative with respect to $K^l_{j,\alpha_1,\dots,\beta}$ yields
		\begin{align*}
			\frac{i}{N}\sum_x\mathrm{tr}_{\alpha_1,\dots,\beta}\left(\mathrm{tr}_\mathrm{rest}\left(M_j^l(s)\right)\left(\sigma^{\alpha_1}\otimes\ \dots\ \otimes\sigma^\beta\right)\right)-2\lambda K^l_{j,\alpha_1,\dots,\beta}(s)=0,
		\end{align*}
	hence,
		\begin{align*}
			K^l_{j,\alpha_1,\dots,\beta}(s)=\frac{i}{2N\lambda}\sum_x\mathrm{tr}_{\alpha_1,\dots,\beta}\left(\mathrm{tr}_\mathrm{rest}\left(M_j^l(s)\right)\left(\sigma^{\alpha_1}\otimes\ \dots\ \otimes\sigma^\beta\right)\right)
		\end{align*}
	This yields the matrix 
		\begin{align*}
			K_j^l(s)&=\sum_{\alpha_1,\dots,\beta}K^l_{j,\alpha_1,\dots,\beta}(s)\left(\sigma^{\alpha_1}\otimes\ \dots\ \otimes\sigma^\beta\right)\\
			&=\frac{i}{2N\lambda}\sum_{\alpha_1,\dots,\beta}\sum_x\mathrm{tr}_{\alpha_1,\dots,\beta}\left(\mathrm{tr}_\mathrm{rest}\left(M_j^l(s)\right)\left(\sigma^{\alpha_1}\otimes\ \dots\ \otimes\sigma^\beta\right)\right)\left(\sigma^{\alpha_1}\otimes\ \dots\ \otimes\sigma^\beta\right)\\
			&=\frac{2^{n_{\alpha_1,\dots,\beta}}i}{2N\lambda}\sum_x\mathrm{tr}_\mathrm{rest}\left(M_j^l(s)\right),
		\end{align*}
	with 
		\begin{align*}
			M_j^l(s)&=\left[U_j^l(s)U_{j-1}^l(s)\dots U_1^1(s)\ \left(\rho_x^\mathrm{in}\otimes\lvert 0\dots 0\rangle_1\langle 0\dots 0\rvert\right) {U_1^1}^\dagger(s)\dots{U_{j-1}^l}^\dagger(s){U_j^l}^\dagger(s),\right.\\
			&\hspace{15pt}\left.{U_{j+1}^l}^\dagger(s)\dots {U_{m_\mathrm{out}}^\mathrm{out}}^\dagger(s)\left(\mathbb{I}_\mathrm{in,hidden}\otimes\lvert\phi^\mathrm{out}_x\rangle\langle\phi^\mathrm{out}_x\rvert\right)U_{m_\mathrm{out}}^\mathrm{out}(s)\dots U_{j+1}^l(s)\right].
		\end{align*}
	As mentioned in the previous subsection, note that $\eta=1/\lambda$ is the learning rate.

\subsection{Efficient Training}

Here, we describe how the channel structure of the feedforward process can be exploited to efficiently train the QNN. Consider a network with $L$ hidden layers and a set of $N$ pairs of training data $\left(\lvert\phi_x^\mathrm{in}\rangle,\lvert\phi_x^\mathrm{out}\rangle\right)$. As described in the previous sections, the general output state of the network at step $s$ is
	\begin{align*}
		\rho_x^\mathrm{out}(s)&=\mathcal{E}_s^{\mathrm{out}}\left(\mathcal{E}_s^{L}\left(\dots\mathcal{E}_s^{2}\left(\mathcal{E}_s^{1}\left(\rho_x^\mathrm{in}\right)\right)\dots\right)\right)
	\end{align*}
with the channel acting on layer $l-1$ and $l$ being
	\begin{align}
	\label{eq:Def-E-Channel}
		\mathcal{E}_s^{l}\left(X^{l-1}\right)&=\mathrm{tr}_{l-1}\left(U_{m_l}^l(s)\dots U_1^l(s)\left(X^{l-1}\otimes\lvert 0\dots 0\rangle_l\langle 0\dots 0\rvert\right){U_1^l}^\dagger(s)\dots{U_{m_l}^l}^\dagger(s)\right),
	\end{align}
where $m_l$ is the number of perceptrons in layer $l$.

This network structure provides a way to compute the derivative of the cost function that is similar to the backpropagation algorithm used in classical machine learning. Consider the cost function
	\begin{equation*}
		C(s)=\frac{1}{N}\sum_{x=1}^N\langle\phi_x^\mathrm{out}\rvert\rho_x^\mathrm{out}(s)\lvert\phi_x^\mathrm{out}\rangle.
	\end{equation*}
To evaluate the derivative of the cost function, we will translate the formula for $\mathrm{d}C(s)/\mathrm{d}s$ (to order $\epsilon$) from \cref{eq:taylordC} to the channel formalism:
	\begin{align*}
		\frac{\mathrm{d}C(s)}{\mathrm{d}s}&=\frac{i}{N}\sum_x \mathrm{tr}\left(\mathbb{I}_\mathrm{in,hidden}\otimes\lvert\phi^\mathrm{out}_x\rangle\langle\phi^\mathrm{out}_x\rvert\left(\left[K_{m_\mathrm{out}}^\mathrm{out}(s), U_{m_\mathrm{out}}^\mathrm{out}(s)\dots U_1^1(s)\left(\rho_x^\mathrm{in}\otimes\lvert 0\dots 0\rangle_\mathrm{hidden,out}\langle 0\dots 0\rvert\right){U_1^1}^\dagger(s)\right.\right.\right.\\
						  &\hspace{15pt}\left.\left.\left.\dots{U_{m_\mathrm{out}}^\mathrm{out}}^\dagger(s)\right]+\dots+U_{m_\mathrm{out}}^\mathrm{out}(s)\dots U_2^1(s) \left[K_1^1(s),U_1^1(s)\left(\rho_x^\mathrm{in}\otimes\lvert 0\dots 0\rangle_\mathrm{hidden,out}\langle 0\dots 0\rvert\right){U_1^1}^\dagger(s)\right]\right.\right.\\&\left.\left.\hspace{15pt}{U_2^1}^\dagger(s)\dots{U_{m_\mathrm{out}}^\mathrm{out}}^\dagger(s)\right)\right)\\
		&=\frac{i}{N}\sum_{x=1}^N \sum_{l=1}^{L+1}\sum_{j=1}^{m_l}\mathrm{tr}\left({U_{1}^{l+1}}^\dagger(s)\dots{U_{m_\mathrm{out}}^\mathrm{out}}^\dagger(s)\left(\mathbb{I}_L\otimes\lvert \phi_x^\mathrm{out}\rangle\langle\phi_x^\mathrm{out}\rvert\right)U_{m_\mathrm{out}}^\mathrm{out}(s)\dots U_{1}^{l+1}(s)\right.\\
		&\hspace{15pt}\left.U_{m_j}^l(s)\dots U_{j+1}^l(s)\left[K_j^l(s),U_j^l(s)\dots U_1^l(s)\left(\rho_x^{l-1}\otimes\lvert 0\dots 0\rangle_l\langle 0\dots 0\rvert\right){U_1^l}^\dagger(s)\dots{U_j^l}^\dagger(s)\right]{U_{j+1}^l}^\dagger(s)\dots{U_{m_j}^l}^\dagger(s)\right)\\
		&=\frac{i}{N}\sum_{x=1}^N \sum_{l=1}^{L+1}\mathrm{tr}\left(\mathcal{F}_s^{l+1}\left(\dots\mathcal{F}_s^{\mathrm{out}}\left(\lvert\phi^\mathrm{out}_x\rangle\langle\phi^\mathrm{out}_x\rvert\right)\dots\right)\right.\\
		&\hspace{15pt}\left.\sum_{j=1}^{m_j}U_{m_j}^l(s)\dots U_{j+1}^l(s)\left[K_j^l(s),U_j^l(s)\dots U_1^l(s)\left(\rho_x^{l-1}\otimes\lvert 0\dots 0\rangle_l\langle 0\dots 0\rvert\right){U_1^l}^\dagger(s)\dots{U_j^l}^\dagger(s)\right]{U_{j+1}^l}^\dagger(s)\dots{U_{m_j}^l}^\dagger(s)\right)\\
		&=\frac{1}{N}\sum_{x=1}^N \sum_{l=1}^{L+1}\mathrm{tr}\left(\sigma_x^{l}(s)\mathcal{D}_s^{l}\left(\rho_x^{l-1}(s)\right)\right),
	\end{align*}
where $\sigma_x^l(s)=\mathcal{F}_s^{l+1}\left(\dots\mathcal{F}_s^{\mathrm{out}}\left(\lvert\phi^\mathrm{out}_x\rangle\langle\phi^\mathrm{out}_x\rvert\right)\dots\right)$ and $\mathcal{D}_s^{l}=\partial\mathcal{E}_s^{l}/\partial s$ the derivative of the corresponding channel, calculated by
	\begin{align*}
		\mathcal{D}_s^{l}\left(X^{l-1}\right)=\sum_{j=1}^{m_j}U_{m_j}^l(s)\dots U_{j+1}^l(s)\left[K_j^l(s),U_j^l(s)\dots U_1^l(s)\left(\rho_x^{l-1}\otimes\lvert 0\dots 0\rangle_l\langle 0\dots 0\rvert\right){U_1^l}^\dagger(s)\dots{U_j^l}^\dagger(s)\right]{U_{j+1}^l}^\dagger(s)\dots{U_{m_j}^l}^\dagger(s)
	\end{align*}
and $\mathcal{F}_s^{l}$ being the adjoint channel of $\mathcal{E}_s^{l}$. The formula for $M_j^l(s)$ in the training algorithm the simplifies to 
	\begin{equation*}
		M_j^l(s)=\left[U_j^l(s)\dots U_1^l(s)\left(\rho_x^{l-1}(s)\otimes \lvert 0\dots 0\rangle_l\langle 0\dots 0\rvert\right){U_1^l}^\dagger(s)\dots{U_j^l}^\dagger (s),{U_{j+1}^l}^\dagger(s)\dots {U_{m_l}^l}^\dagger(s)\left(\mathbb{I}_l\otimes \sigma_x^l(s)\right)U_{m_l}^l(s)\dots U_{j+1}^l(s)\right].
	\end{equation*}
	It will we be useful for the implementation of the network to have an explicit expression of the adjoint channel $\mathcal{F}_s^{l}$.
	In order to obtain this we write the channel $\mathcal{E}_s^{l}$ in its Kraus representation, which is for any operator $X^{l-1}$ on the $(l-1)$th layer
	\begin{equation*}
	\mathcal{E}_s^{l}(X^{l-1}) = \sum_\alpha A_\alpha X^{l-1}A^\dag_\alpha.
	\end{equation*}
	Here we have omitted the indices $s$ and $l$ for the Kraus operators $A_\alpha$ to make the notation clearer. Note that each of the Kraus operators $A_\alpha$ is a map from the $(l-1)$th layer consisting of $m_{l-1}$ qubits to the $l$th layer consisting of $m_l$ qubits.
	The adjoint channel $\mathcal{F}_s^{l}$ is then by definition given by
	\begin{equation}
	\label{eq:defFChannel}
	\mathcal{F}_s^{l}(X^{l}) = \sum_\alpha A^\dag_\alpha X^{l}A_\alpha,
	\end{equation}
	for any operator $X^l$ on the $l$th layer.
	
	We are now seeking for an explicit formula of the Kraus operators $A_\alpha$. Let $\{\ket{\alpha}\}_\alpha$ be an orthonormal basis in the $(l-1)$th layer. Moreover, let $\ket{m},\ket{n}$ be any vectors in the $(l-1)$th layer and $\ket{i},\ket{j}$ any vectors in the $l$th layer. Then the action of $\mathcal{E}_s^{l}$ can be calculated using \eqref{eq:Def-E-Channel} and the shorthand notation $U^l(s) = U^l_{m_l}(s)\dots U^l_1(s)$ for the whole unitary of the layer $l$, which gives
	\begin{align*}
	\big\langle i\big|\, \mathcal{E}_s^{l}\left(\ket{m}\bra{n}\right)\big| j\big\rangle &=\Big\langle i\Big|\,\mathrm{tr}_{l-1}\left(U^l(s)\left(\ket{m}\bra{n}\otimes\lvert 0\dots 0\rangle_l\langle 0\dots 0\rvert_l\right){U^l}^\dagger(s)\right)\Big| j \Big\rangle  \\
	&= \sum_\alpha\big\langle \alpha,i\big|\,U^l(s)\left(\ket{m}\bra{n}\otimes\lvert 0\dots 0\rangle_l\langle 0\dots 0\rvert_l\right){U^l}^\dagger(s)\big|\alpha, j \big\rangle \\
	&= \sum_\alpha\big\langle \alpha,i\big|\,U^l(s)\big| m, 0\dots 0\big\rangle\big\langle n, 0\dots 0\big|{U^l}^\dagger(s)\big|\alpha, j \big\rangle.
	\end{align*}
	Therefore, defining $A_\alpha$ via $\bra{i}A_\alpha\ket{m} = \big\langle \alpha,i\big|\,U^l(s)\big| m, 0\dots 0\big\rangle$ this gives a set Kraus operators for $\mathcal{E}_s^{l}$. Using this definition and \eqref{eq:defFChannel} we obtain
	\begin{align*}
	\bra{m}\mathcal{F}_s^{l}(\ket{i}\bra{j})\ket{n} &= \sum_\alpha \bra{m}A^\dag_\alpha \ket{i}\bra{j}A_\alpha\ket{n} = \sum_\alpha\big\langle m,0\dots 0\big|\,{U^l}^\dag(s)\big| \alpha, i\big\rangle\big\langle \alpha, j \big|U^l(s)\big|n,0 \dots 0 \big\rangle \\
	&= \big\langle m,0\dots 0\big|\,{U^l}^\dag(s)\left(\mathbb{I}_{l-1}\otimes\ket{i}\bra{j}\right)U^l(s)\big|n,0 \dots 0 \big\rangle \\
	&= \Big\langle m\Big|\,\mathrm{tr}_{l}\left(\mathbb{I}_{l-1}\otimes|0\dots 0\rangle_l\langle0\dots 0|_l{U^l}^\dag(s)\left(\mathbb{I}_{l-1}\otimes\ket{i}\bra{j}\right)U^l(s)\right)\Big|n\Big\rangle.
	\end{align*}
	From this we already know the action of $\mathcal{F}_s^{l}$ on a general operator $X^l$, which is
	\begin{equation*}
	\mathcal{F}_s^{l}(X^l) = \mathrm{tr}_{l}\left(\mathbb{I}_{l-1}\otimes|0\dots 0\rangle_l\langle0\dots 0|_l{U^l}^\dag(s)\left(\mathbb{I}_{l-1}\otimes X^l\right)U^l(s)\right).
	\end{equation*}

\section{Estimating the optimal cost function for learning an unknown unitary}

In this section we derive estimates for the typical value of the cost function when learning an unknown unitary $V$ acting on $D$-dimensional qudit. We focus on the setting where we have access to $N$ training pairs $\left(\lvert\phi_x\rangle,V\lvert\phi_x\rangle\right)$, $x=1,2,\dots, N$, where $|\phi_x\rangle$ have been chosen uniformly at random according to the Haar measure induced on state space \footnote{The number $N$ of training pairs may exceed the dimension $D$ of the input space.}. We use the first $n$ ($< D$) pairs to train the network and then we investigate how well the network was trained by evaluating the cost function for all of the $N$ pairs.

With probability $1$ any subset of $D$ of the states $|\phi_x\rangle$ will be linearly independent. Thus the first $n<D$ states $\lvert\phi_x\rangle$ span, with probability $1$ an $n$-dimensional subspace $\mathcal{K}\subset \mathcal{H}\cong \mathbb{C}^D$ which is unitarily mapped by $V$ onto an $n$-dimensional subspace $\mathcal{L}$:
	\begin{align*}
		\mathcal{K}&=\mathrm{span}\{\lvert\phi_x\rangle,x=1,2,\dots,n\},\\
		\mathcal{L}&=\mathrm{span}\{V\lvert\phi_x\rangle,x=1,2,\dots,n\}.
	\end{align*}
We actually also consider a second scenario in the sequel, namely, where we generate $n<D$ random \emph{orthogonal} input states $|\phi_x\rangle$ for the first $n$ training pairs. In this case there is a quantitative difference in the performance of the quantum neural network due to ambiguity with phases.

Suppose our network is expressive enough that we can use it to represent the \emph{best} unitary $W$ for the available data. In the case where the training data is chosen completely at random this implies
\begin{align*}
		W\lvert\phi_x\rangle=V\lvert\phi_x\rangle,
	\end{align*}
for $x = 1,2, \ldots, n$. In the case the initial training data consists entirely of orthogonal states then all we can say is that 
	\begin{align*}
		W\lvert\phi_x\rangle=e^{i\theta_x}V\lvert\phi_x\rangle,
	\end{align*}
because the most we can infer from maximising the cost function is that $W$ acts like $V$ (up to a phase that depends on the state) on the states in $\mathcal{K}$, but we have no further information about how $V$ acts on the rest of the space (apart from the fact that it maps $\mathcal{K}^\perp$ to $\mathcal{L}^\perp$). Hence, in both cases, the learned unitary $W$ can be written as
	\begin{align*}
		W=\sum_{x=1}^n e^{i\theta_x}V\lvert\phi_x\rangle\langle\phi_x\rvert+\sum_{x=n+1}^N W^\perp \lvert\phi_x\rangle\langle\phi_x\rvert,
	\end{align*}
	where the phases $\theta_x =0$ when the input data is not orthogonal and $\theta_x$ are undertermined when the initial data is orthogonal.
The corresponding cost function for all of the training data $\left(\lvert\phi_x\rangle,V\lvert\phi_x\rangle\right)$, $x=1,2,\dots, N$, is then
	\begin{align*}
		C&=\frac{1}{N}\sum_{x=1}^n\langle\phi_x\rvert V^\dagger W\lvert\phi_x\rangle\langle\phi_x\rvert W^\dagger V\lvert\phi_x\rangle+\frac{1}{N}\sum_{x=n+1}^N\langle\phi_x\rvert V^\dagger W\lvert\phi_x\rangle\langle\phi_x\rvert W^\dagger V\lvert\phi_x\rangle\\
		&=\frac{n}{N}+\frac{1}{N}\sum_{x=n+1}^N\mathrm{tr}\Big(\underbrace{V\lvert\phi_x\rangle\langle\phi_x\rvert V^\dagger}_{\equiv Q_x} W \underbrace{\lvert\phi_x\rangle\langle\phi_x\rvert}_{\equiv P_x} W^\dagger\Big)\\
		&=\frac{n}{N}+\frac{1}{N}\sum_{x=n+1}^N\mathrm{tr}\Big(Q_x WP_xW^\dagger\Big).
	\end{align*}
Our input training states $|\phi_x\rangle$, $x = 1,2,\ldots, N$ were chosen at random. To understand the average case behaviour of the cost function we now take an expectation value with respect to this measure:
	\begin{equation*}
		\overline{C} \equiv \mathbb{E}\left[C\right]=\frac{n}{N}+\frac{1}{N}\sum_{x=n+1}^N\mathbb{E}\left[\mathrm{tr}\left(Q_xWP_xW^\dagger\right)\right]
	\end{equation*}
To evaluate this expectation value we exploit the identity
\begin{equation*}
	\int d|\phi\rangle \, |\phi\rangle\langle\phi|\otimes |\phi\rangle\langle\phi| = \frac{2}{D(D+1)}P_{\text{sym}} = \frac1{D(D+1)}\left(\mathbb{I}\otimes \mathbb{I} + \textsc{swap}\right).
\end{equation*}
Accordingly
\begin{align*}
	\mathbb{E}\left[\langle\phi_x\rvert V^\dagger W\lvert\phi_x\rangle\langle\phi_x\rvert W^\dagger V\lvert\phi_x\rangle\right] &= \frac1{D(D+1)}\sum_{j,k=1}^D \text{tr}\left(V^\dag W |j\rangle\langle j| W^\dag V |k\rangle\langle k|\right) + \text{tr}\left(V^\dag W |j\rangle\langle k| W^\dag V |k\rangle\langle j|\right)\\
	&= \frac1{D(D+1)}\left({D} +\left|\text{tr}(V^\dag W )\right|^2\right).
\end{align*}

The average value of the full cost function is therefore
	\begin{equation*}
		\overline{C} \equiv \mathbb{E}\left[C\right]=\frac{n}{N}+\frac{N-n}{ND(D+1)}\left({D} + \left|\text{tr}(V^\dag W )\right|^2\right).
	\end{equation*}

Our next step is to estimate the quantity $\left|\text{tr}(V^\dag W )\right|^2$. Since $W$ is identical to $V$ up to a phase on $\mathcal{K}$ we have that
\begin{equation*}
	\left|\text{tr}(V^\dag W )\right|^2 = \left|\sum_{x=1}^{n} e^{i\theta_x}+\text{tr}(X^\perp)\right|^2,
\end{equation*}
where $X^\perp = V^\perp (W^\perp)^\dag$. The unitary $X^\perp$ acting on the $(D-n)$-dimensional subspace $\mathcal{K}^\perp$ is completely unknown. Therefore the best we can do is simply guess $X^\perp$ and the $n$ phases $e^{i\theta_x}$ uniformly at random according to Haar measure. Thus, to estimate the value of the cost function after this guessing we take a \emph{second} expectation value, this time over $X^\perp$ and the phases:
\begin{equation*}
	\mathbb{E}_{\theta_x,X^\perp}\left[\left|\sum_{x=1}^{n} e^{i\theta_x}+\text{tr}(X^\perp)\right|^2\right] = \min\{n+1,D\},
\end{equation*}
where we've used the result that the second moment of $\text{tr}(X)$ when averaged over the Haar measure is equal to $1$ \cite{DS1994,DE2001}.

Putting this together we get that, on average, the cost function for the best possible guess for $W$, given the training data, should behave as 
	\begin{equation*}
		C\sim \frac{n}{N}+\frac{N-n}{ND(D+1)}\left(D+\min\{n+1,D\}\right).
	\end{equation*}
	In the case that the initial training data was comprised of random states we obtain instead
\begin{align}\label{bla}
 C\sim \frac{n}{N}+\frac{N-n}{ND(D+1)}\left(D+\min\{n^2+1,D^2\}\right).
\end{align}

% !TeX spellcheck = en_GB

\section{Results}

To test how well the quantum neural network performs, we have simulated it with MATLAB and with Mathematica (the code is available at \url{https://github.com/R8monaW/DeepQNN}) and used it for different tasks.

\subsection{Generalisation}

The first task we consider aims at understanding how well the QNN is able to generalise, which means that the number of training pairs we use is fewer than the Hilbert space dimension. We have studied the performance of the QNN for different network architectures and different choices for the parameters and the Hilbert space dimension. In all plots, the violet points are the estimated values of the cost function according to~\eqref{bla} and the orange points are the numerical values. The results are depicted below.

\begin{figure}[H]
	\centering
\begin{subfigure}{0.48\textwidth}
\raisebox{2.2mm}{
	\begin{tikzpicture}[yscale=5,xscale=2]
		\def\xmin{0}
		\def\xmax{4}
		\def\ymin{0}
		\def\ymax{1}
		\draw[->, thin, draw=gray] (\xmin,\ymin)--(\xmax,\ymin) node [below left] {\scriptsize Number of training pairs}; 
		\draw[->, thin, draw=gray] (\xmin,\ymin)--(\xmin,\ymax)node [above right] {\scriptsize Cost for test pairs};  
		\foreach \x in {1,2}
		\node at (\x, \ymin) [below] {\scriptsize\x};
		\foreach \y in {.1,.2,.3,.4,.5,.6,.7,.8,.9}
		\node at (\xmin,\y) [left] {\scriptsize\y};
		%put in the next line data after "in"
		\foreach \Point in {(1, 0.3306061809796075), (2, 0.548298996171964 ), (3, 0.832766574049491 ), (4, 0.9855403146185742)}{
			\node[color2] at \Point {\textbullet};
		}
		\foreach \Point in {(1, 0.37), (2, 0.56), (3, 0.79), (4, 1)}{
			\node[color1] at \Point {\textbullet};
		}
		\draw [xstep=1,ystep=.1,gray, dotted]  (4,0) grid (0,1);
\end{tikzpicture}}
\caption{We have trained a \begin{tikzpicture}[yscale=0.7,xscale=0.6, baseline]
\node(1) [circle,draw,inner sep=0pt,minimum size=4.5pt] at (-1,0) {};
\node(2) [circle,draw,inner sep=0pt,minimum size=4.5pt] at (-1,0.3) {};
\node(4) [circle,draw,inner sep=0pt,minimum size=4.5pt] at (0,0) {};
\node(5) [circle,draw,inner sep=0pt,minimum size=4.5pt] at (0,0.3) {};
\draw (1)--(4);
\draw (2)--(4);
\draw (1)--(5);
\draw (2)--(5);
\end{tikzpicture}  network for $2000$ rounds each time with parameters $\eta=1$, $\epsilon=0.1$ and averaged the cost function over $20$ runs. We used $10$ pairs in the test set.}
\label{fig:resultsaA}
\end{subfigure}
\hfill
\begin{subfigure}{0.48\textwidth}
	\begin{tikzpicture}[yscale=5,xscale=1]
	\def\xmin{0}
	\def\xmax{8}
	\def\ymin{0}
	\def\ymax{1}
	\draw[->, thin, draw=gray] (\xmin,\ymin)--(\xmax,\ymin) node [below left] {\scriptsize Number of training pairs}; 
	\draw[->, thin, draw=gray] (\xmin,\ymin)--(\xmin,\ymax)node [above right] {\scriptsize Cost for test pairs};  
	\foreach \x in {1,...,4}
	\node at (\x, \ymin) [below] {\scriptsize\x};
	\foreach \y in {.1,.2,.3,.4,.5,.6,.7,.8,.9}
	\node at (\xmin,\y) [left] {\scriptsize\y};
	%put in the next line data after "in"
	\foreach \Point in {(1, 0.22019666130478677),(2, 0.34456619098462443),(3, 0.4672191059180644),(4, 0.6153667918789689),(5, 0.7393789506540993),(6, 0.8569043801723268),(7, 0.9481239799787362),(8, 0.9854768291157662)}{
		\node[color2] at \Point {\textbullet};
	}
	\foreach \Point in {(1, 0.225), (2, 0.344444), (3, 0.475), (4, 0.608333), (5, 0.736111), (6, 0.85), (7, 0.941667), (8, 1)}{
		\node[color1] at \Point {\textbullet};
	}
	\draw [xstep=1,ystep=.1,gray, dotted]  (8,0) grid (0,1);
	\end{tikzpicture}
\caption{We have trained a \begin{tikzpicture}[yscale=0.7,xscale=0.6, baseline]
		\node(1) [circle,draw,inner sep=0pt,minimum size=4.5pt] at (-1,0) {};
		\node(2) [circle,draw,inner sep=0pt,minimum size=4.5pt] at (-1,0.3) {};
		\node(3) [circle,draw,inner sep=0pt,minimum size=4.5pt] at (-1,0.6) {};
		\node(4) [circle,draw,inner sep=0pt,minimum size=4.5pt] at (0,0) {};
		\node(5) [circle,draw,inner sep=0pt,minimum size=4.5pt] at (0,0.3) {};
		\node(6) [circle,draw,inner sep=0pt,minimum size=4.5pt] at (0,0.6) {};
		\node(7) [circle,draw,inner sep=0pt,minimum size=4.5pt] at (1,0) {};
		\node(8) [circle,draw,inner sep=0pt,minimum size=4.5pt] at (1,0.3) {};
		\node(9) [circle,draw,inner sep=0pt,minimum size=4.5pt] at (1,0.6) {};
		\draw (1)--(4);
		\draw (2)--(4);
		\draw (3)--(4);
		\draw (1)--(5);
		\draw (2)--(5);
		\draw (3)--(5);
		\draw (1)--(6);
		\draw (2)--(6);
		\draw (3)--(6);
		\draw (4)--(7);
		\draw (5)--(7);
		\draw (6)--(7);
		\draw (4)--(8);
		\draw (5)--(8);
		\draw (6)--(8);
		\draw (4)--(9);
		\draw (5)--(9);
		\draw (6)--(9);
	\end{tikzpicture}
 network with $\epsilon=0.1$, $\eta=2/3$ for $1000$ rounds and averaged the cost function over $20$ runs, using $10$ pairs in the test set.}
\label{fig:resultsbA}
\end{subfigure}
\caption{Numerical results for the generalisation task.}
\label{fig:resultsA}
\end{figure}
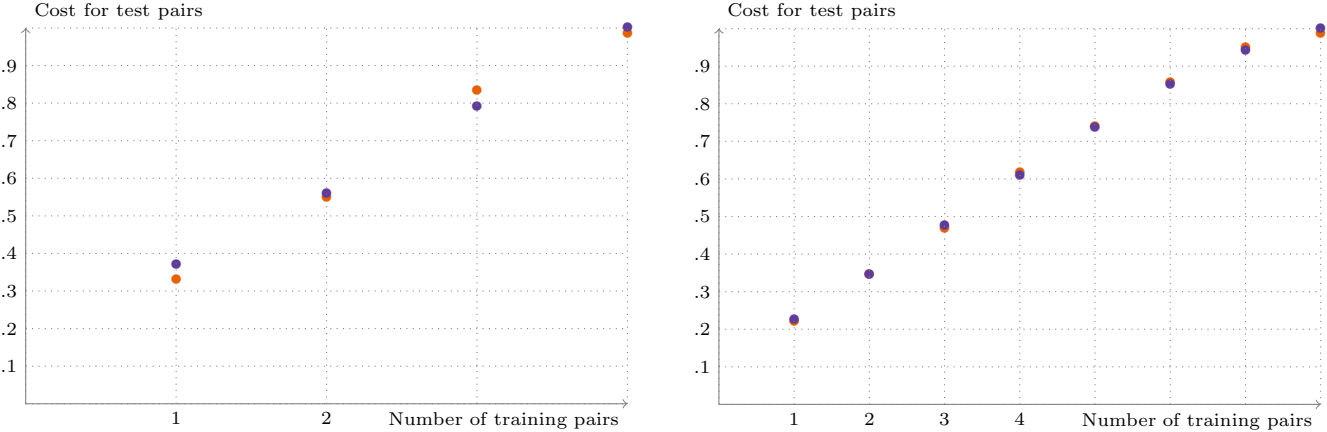
\subsection{Robustness to Noisy Data}

The second task we studied was about understanding the robustness of the QNN to noisy training data. We have generated $N$ good training pairs and then corrupted $n$ of them by replacing them with random pairs. It is chosen randomly which of the good pairs are corrupted. In all plots, the number on the $x$-axis indicates how many of the good training pairs were replaced by a pair of random states and the cost function is evaluated for all good test  pairs. Again, we have studied different network architectures, parameters and dimensions, as depicted in the figures below.

\begin{figure}[H]
	\centering
\begin{subfigure}{0.48\textwidth}
\raisebox{2.2mm}{
	\begin{tikzpicture}[yscale=5,xscale=.08]
	\def\xmin{0}
	\def\xmax{100}
	\def\ymin{0}
	\def\ymax{1}
	\draw[->, thin, draw=gray] (\xmin,\ymin)--(\xmax,\ymin) node [below left] {\scriptsize Number of noisy pairs}; 
	\draw[->, thin, draw=gray] (\xmin,\ymin)--(\xmin,\ymax) node [above right] {\scriptsize Cost for good test pairs};  
	\foreach \x in {10,20,...,60}
	\node at (\x, \ymin) [below] {\scriptsize\x};
	\foreach \y in {.1,.2,.3,.4,.5,.6,.7,.8,.9}
	\node at (\xmin,\y) [left] {\scriptsize\y};
	%put in the next line data after "in"
	\foreach \Point in {(0, 1.), (5, 0.999825), (10, 0.999544), (15, 0.999371), (20, 0.999332), (25, 0.997853), (30, 0.997012), (35, 0.99648), (40, 0.99765), (45, 0.995095), (50, 0.988275), (55, 0.98431), (60, 0.979979), (65, 0.976593), (70, 0.961271), (75, 0.922404), (80, 0.792853), (85, 0.617134), (90, 0.336601), (95, 0.303163), (100, 0.248403)}{
		\node[color2] at \Point {\textbullet};
	}
	\draw [xstep=10,ystep=.1,gray, dotted]  (100,0) grid (0,1);
	\end{tikzpicture}}
	\caption{We have trained a \begin{tikzpicture}[yscale=0.7,xscale=0.6, baseline]
\node(1) [circle,draw,inner sep=0pt,minimum size=4.5pt] at (-1,0) {};
\node(2) [circle,draw,inner sep=0pt,minimum size=4.5pt] at (-1,0.3) {};
\node(4) [circle,draw,inner sep=0pt,minimum size=4.5pt] at (0,0) {};
\node(5) [circle,draw,inner sep=0pt,minimum size=4.5pt] at (0,0.3) {};
\draw (1)--(4);
\draw (2)--(4);
\draw (1)--(5);
\draw (2)--(5);
\end{tikzpicture} network with $\epsilon=0.1$, $\eta=1$ for 500 rounds with $100$ training pairs.}
	\label{fig:resultB1}
\end{subfigure}
\hfill
\begin{subfigure}{0.48\textwidth}
	\centering
	\begin{tikzpicture}[yscale=5,xscale=.08]
	\def\xmin{0}
	\def\xmax{100}
	\def\ymin{0}
	\def\ymax{1}
	\draw[->, thin, draw=gray] (\xmin,\ymin)--(\xmax,\ymin) node [below left] {\scriptsize Number of noisy pairs}; 
	\draw[->, thin, draw=gray] (\xmin,\ymin)--(\xmin,\ymax) node [above right] {\scriptsize Cost for good test pairs};  
	\foreach \x in {10,20,...,60}
	\node at (\x, \ymin) [below] {\scriptsize\x};
	\foreach \y in {.1,.2,.3,.4,.5,.6,.7,.8,.9}
	\node at (\xmin,\y) [left] {\scriptsize\y};
	%put in the next line data after "in"
	\foreach \Point in {(0, 0.9999988812037631), (5, 0.9998675362562289), (10, 0.9991545815565356), (15, 0.9980864880626191), (20, 0.9981773438490551), (25, 0.9973477437970019), (30, 0.996208256972089), (35, 0.9948700240610997), (40, 0.9947395171844108), (45, 0.9924308541328096), (50, 0.9865230944649602), (55, 0.9883202771721912), (60, 0.9733380112449577), (65, 0.9607046692132285), (70, 0.9579410995168592), (75, 0.8729391963495594), (80, 0.7336290899974561), (85, 0.5779847247543635), (90, 0.4171436560766765), (95, 0.2559509395527556), (100, 0.23817485576276787)}{
		\node[color2] at \Point {\textbullet};
	}
	\draw [xstep=10,ystep=.1,gray, dotted]  (100,0) grid (0,1);
	\end{tikzpicture}
	\caption{We have trained a \begin{tikzpicture}[yscale=0.7,xscale=0.6, baseline]
		\node(1) [circle,draw,inner sep=0pt,minimum size=4.5pt] at (-1,.15) {};
		\node(2) [circle,draw,inner sep=0pt,minimum size=4.5pt] at (-1,0.45) {};
		\node(4) [circle,draw,inner sep=0pt,minimum size=4.5pt] at (0,0) {};
		\node(5) [circle,draw,inner sep=0pt,minimum size=4.5pt] at (0,0.3) {};
		\node(6) [circle,draw,inner sep=0pt,minimum size=4.5pt] at (0,0.6) {};
		\node(7) [circle,draw,inner sep=0pt,minimum size=4.5pt] at (1,.15) {};
		\node(8) [circle,draw,inner sep=0pt,minimum size=4.5pt] at (1,0.45) {};
		\draw (1)--(4);
		\draw (2)--(4);
		\draw (1)--(5);
		\draw (2)--(5);
		\draw (1)--(6);
		\draw (2)--(6);
		\draw (4)--(7);
		\draw (5)--(7);
		\draw (6)--(7);
		\draw (4)--(8);
		\draw (5)--(8);
		\draw (6)--(8);
	\end{tikzpicture} network with $\epsilon=0.1$, $\eta=1$ for $300$ rounds with $100$ training pairs.}
	\label{fig:resultB2}
\end{subfigure}
\caption{Numerical results for estimating the robustness of the QNN to noisy data.}
\end{figure}
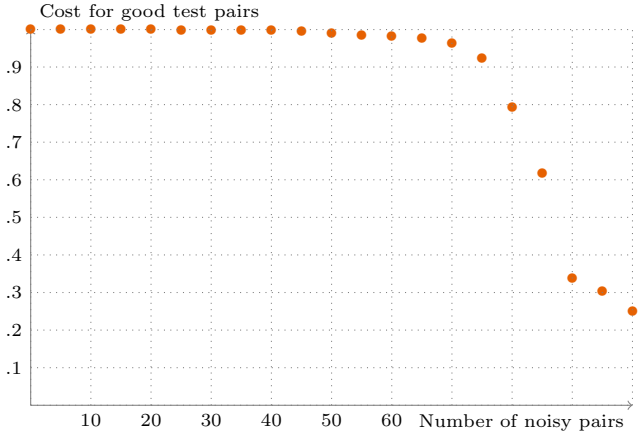
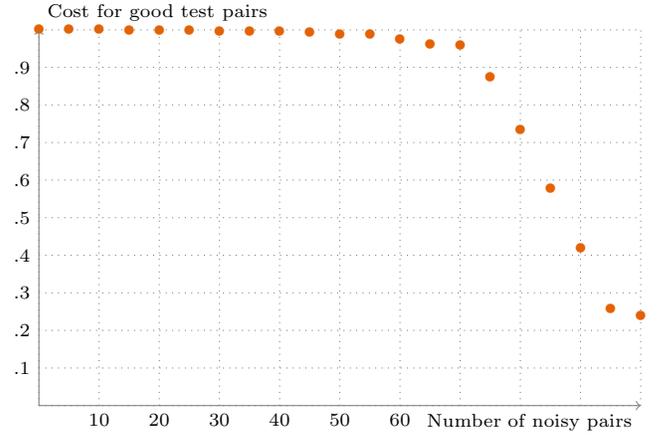

\subsection{Deep Neural Networks}

Beside the previous tasks, we have also studied how well deep neural networks train in the classical simulation.
\begin{figure}[H]
\input{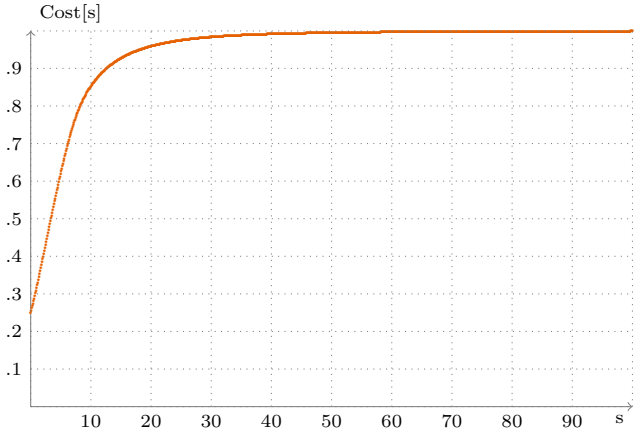}\hfill\input{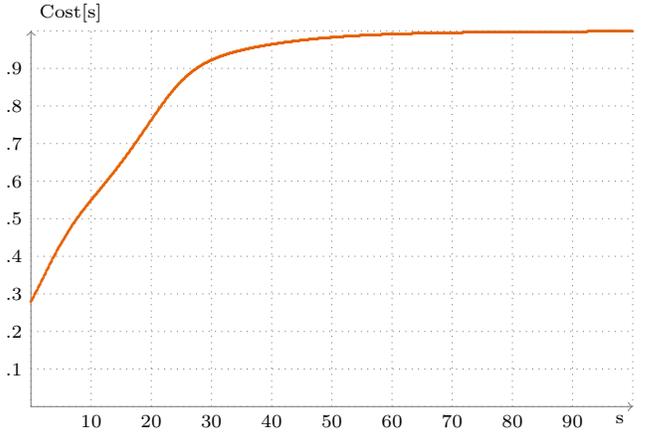}
\caption{Numerical results for deep neural networks.}
\end{figure}

\section{Quantum algorithm for quantum training of the neural network}
\label{sec:QuantumNNopt}
In this section we explain how our algorithm can be implemented on a quantum computer. To begin we want to clarify what operations a quantum computer is assumed to be able to do in our case:
\begin{itemize}
\item[1.] Partial trace.
\item[2.] Initialize a qubit in $\ket{0}$ state.
\item[3.] Apply $\cnot$, $T$, $H$ (and therefore perceptrons \footnote{The Solovay-Kitaev theorem says that any 2-qubit unitary can be built out of $O(\log^c(\frac{1}{\epsilon}))$ gates, where $\epsilon$ is the accuracy \cite{Nielsen2010}}) easily.
\item[4.] Measuring in computational basis.
\end{itemize}
From now on we have two tasks. We need to compute the cost function as well as work out the derivative of the cost function on a quantum computer. We label/describe these two tasks as subroutine 1 and subroutine 2, respectively. 

\subsection{Subroutine 1}
In this subroutine we use the ``SWAP trick" to estimate the fidelity of a pure state $\ket{\phi}$ with a mixed state $\rho$. Our input is the state $\ket{\phi}$ in a register of $m$ qubits and $\rho$ in another register of $m$ qubits. In total we have $2m$ qubits, however, we require an additional ancillary qubit for the following process. We estimate $F(\ket{\phi},\rho)=\braket{\phi|\rho|\phi}$ as a probability exploiting the following quantum circuit.
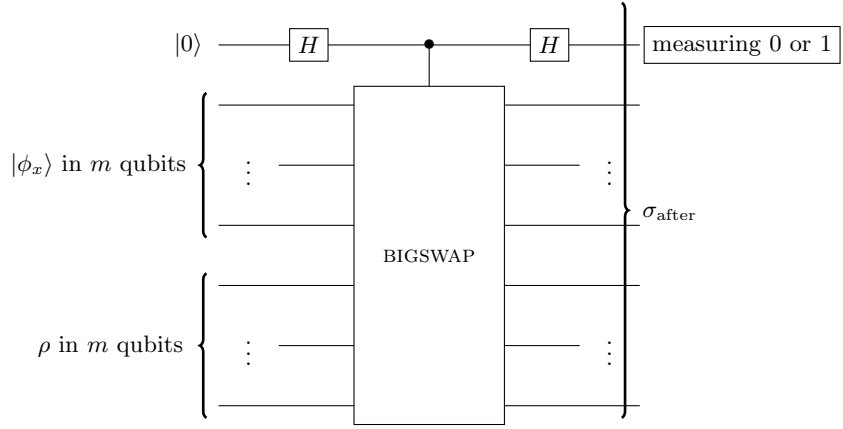
\begin{figure}[H]
\centering
\begin{tikzpicture}[scale=.8, decoration={brace}]
\draw(0,6)--(7,6);	
\draw(0,5)--(7,5);	
\draw(1,4)--(6,4);	
\draw(0,3)--(7,3);	
\draw(0,2)--(7,2);	
\draw(1,1)--(6,1);	
\draw(0,0)--(7,0);	
\draw (3.5,2.5)--(3.5,6);
\node (a) at (3.5,6) {\textbullet};
\node (a) at (.5,1) {$\vdots$};
\node (b) at (6.5,1) {$\vdots$};
\node (c) at (.5,4) {$\vdots$};
\node (d) at (6.5,4) {$\vdots$};
\node[draw,minimum height=4.5cm,minimum width=2cm,fill=white] (e) at (3.5,2.5) {$\bigswap$};
\node[draw,fill=white] (f) at (1.5,6) {$H$};
\node[draw,fill=white] (g) at (5.5,6) {$H$};
\node[draw,fill=white] (h) at (8.7,6) {measuring 0 or 1};
\node (i) at (-.5,6) {$\ket{0}$};
\node (i) at (-2.0,4) {$\ket{\phi_x}$ in $m$ qubits};
\node (i) at (-1.8,1) {$\rho$ in $m$ qubits};
\node (i) at (7.5,3.2) {$\sigma_{\text{after}}$};
\draw [decorate,line width=1pt] (-.2,-.2) -- (-.2,2.2);
\draw [decorate,line width=1pt] (-.2,2.8) -- (-.2,5.2);
\draw [decorate,line width=1pt] (6.7,6.7) -- (6.7,-.2);
\end{tikzpicture}
\captionof{figure}{Quantum circuit for computing the cost function.}
\end{figure}

To explain our subroutine we assume $m=1$ for simplicity.

\begin{itemize}
\item[a.] \textbf{Initialization:}
We initialize the $2m+1$ qubits in the state
\begin{align*}
&\ket{0}\bra{0}\otimes\ket{\phi}\bra{\phi}\otimes\rho\text{.}
\intertext{
\item[b.] \textbf{Hadamard:}
In the next step we apply the Hadamard gate and end up with the state}
&\frac{1}{2}(\ket{0}+\ket{1})(\bra{0}+\bra{1})\otimes\ket{\phi}\bra{\phi}\otimes\rho\text{.}
\intertext{
\item[c.] \textbf{CSWAP:} 
We use $\cswap:=\ket{0}\bra{0}\otimes\mathbbm{1}+\ket{1}\bra{1}\otimes \swap$ and the result is:}
	&\cswap^\dagger\left(\frac{1}{2}(\ket{0}+\ket{1})(\ket{0}+\ket{1})\otimes\ket{\phi}\bra{\phi}\otimes\rho\right)\cswap\\   =&\frac{1}{2}\ket{0}\bra{0}\otimes\ket{\phi}\bra{\phi}\otimes\rho+\frac{1}{2}\ket{1}\bra{0}\left(\swap(\ket{\phi}\bra{\phi}\otimes\rho)\right)+\frac{1}{2}\ket{0}\bra{1}\left((\ket{\phi}\bra{\phi}\otimes\rho)\swap\right)\\&+\frac{1}{2}\ket{1}\bra{1}\left(\swap(\ket{\phi}\bra{\phi}\otimes\rho)\swap\right).
\intertext{
\item[d.] \textbf{Hadamard:}
After applying the Hadamard gate a second time we have the following expression:}	
\sigma_{\text{after}}=&\frac{1}{4}(\ket{0}+\ket{1})(\bra{0}+\bra{1})\otimes(\ket{\phi}\bra{\phi}\otimes\rho)+\frac{1}{4}(\ket{0}-\ket{1})(\bra{0}+\bra{1})\otimes\left(\swap(\ket{\phi}\bra{\phi}\otimes\rho)\right)\\&+\frac{1}{4}(\ket{0}+\ket{1})(\bra{0}-\bra{1})\otimes\left((\ket{\phi}\bra{\phi}\otimes\rho)\swap\right)+\frac{1}{4}(\ket{0}-\ket{1})(\bra{0}-\bra{1})\otimes\left(\swap(\ket{\phi}\bra{\phi}\otimes\rho)\swap\right).
\intertext{
\item[e.] \textbf{Measuring:}
In this last step we measure the first control qubit and get $0$ with probability $p_0$.}
p_0=&\tr(\ket{0}\bra{0}\otimes\mathbbm{1}\otimes\mathbbm{1}\times\sigma_{\text{after}})\\
=&\frac{1}{4}\tr(\ket{0}\bra{0}\big((\ket{0}+\ket{1})(\bra{0}+\bra{1})\big))\tr(\ket{\phi}\ket{\phi}\otimes\rho)+\frac{1}{4}\tr(\ket{0}\bra{0}\big((\ket{0}-\ket{1})(\bra{0}+\bra{1})\big))\tr\left(\swap(\ket{\phi}\bra{\phi}\otimes\rho)\right)\\&+\frac{1}{4}\tr(\ket{0}\bra{0}\big((\ket{0}+\ket{1})(\bra{0}-\bra{1}))\tr\left((\ket{\phi}\bra{\phi}\otimes\rho)\swap\right)\\&+\frac{1}{4}\tr(\ket{0}\bra{0}\big((\ket{0}-\ket{1})(\bra{0}-\bra{1}))\tr\left(\swap(\ket{\phi}\bra{\phi}\otimes\rho)\swap\right)
\\
=&\frac{1}{4}+\frac{1}{4}\tr(\swap\ket{\phi}\bra{\phi}\otimes\rho)+\frac{1}{4}\tr(\ket{\phi}\bra{\phi}\otimes\rho\swap)+\frac{1}{4}\\
=&\frac{1}{2}+\frac{1}{2}\tr(\swap\ket{\phi}\bra{\phi}\otimes\rho).
\intertext{
Using the definition $\swap=\sum_{j,k=1}^2\ket{jk}\bra{kj}$ we obtain:}
=&\frac{1}{2}+\frac{1}{2}\sum_{j,k}\tr(\ket{jk}\bra{kj}(\ket{\phi}\bra{\phi}\otimes\rho)\\
	     =&\frac{1}{2}+\frac{1}{2}\sum_{j,k}\braket{k|\phi}\braket{j|\rho|k}\braket{\phi|j}\\
	     =&\frac{1}{2}+\frac{1}{2}\sum_{j,k}\braket{\phi|j}\braket{j|\rho|k}\braket{k|\phi}\\
	     =&\frac{1}{2}+\frac{1}{2}F\left(\ket{\phi},\rho\right).
\end{align*}
\end{itemize}
At this point we encounter quantum projective noise, i.e., we get $0$ or $1$ randomly and need to repeat this measurement $N$ times to reduce the fluctuations arising from the bionomial probability distribution.  
We get 
\begin{align*}
	\frac{\#0\text{s}}{N}&=p_0+\delta p_0\\
	\frac{\#1\text{s}}{N}&=p_1+\delta p_1
\end{align*}
with fluctuations $\delta p_i=\sqrt{\frac{p_i(p_i-1)}{N}}\approx\frac{p_i}{\sqrt{N}}$.
Our resource usage so far amounts to:
\begin{itemize}
	\item $2N$ Hadamards,
	\item $N$ copies of $\ket{\phi}$,
	\item $N$ copies of $\rho$, and
	\item $N$ $\cswap$s.
\end{itemize}
In addition to that we need $m$ qubits for the operation 
\begin{equation*}
\cswap=\ket{0}\bra{0}\otimes\mathbbm{1}\otimes\mathbbm{1}+\ket{1}\bra{1}\otimes\bigswap,
\end{equation*}
where
\begin{equation*}
	\bigswap=\sum_{j_1,\hdots j_m;k_1,\hdots,k_m} \ket{j_1,\hdots,j_m;k_1,\hdots,k_m,}\bra{k_1,\hdots,k_m;j_1,\hdots,j_m}\text{.}
\end{equation*}

\begin{figure}[H]
\centering
\begin{subfigure}{.25\textwidth}
\centering
\begin{tikzpicture}[decoration={markings,mark=at position .5 with {\arrow[scale=1,thick]{>}}}] 
\begin{knot}[flip crossing/.list={1,2}]
\strand (0,0)--(1,0)--(2,1)--(3,1);
\strand (0,1)--(1,1)--(2,0)--(3,0);\end{knot}
\end{tikzpicture}
\captionof{figure}{$\swap$ gate.}
\end{subfigure}
\begin{subfigure}{.25\textwidth}
\centering
\begin{tikzpicture}[scale=.4] 
\draw [decorate,decoration={brace,amplitude=10pt},xshift=-4pt,yshift=0pt]
(0,6) -- (0,10) node [black,midway,xshift=-0.6cm] 
{\footnotesize $m$};
\node at (.5,1)   (a) {$\vdots$};
\node at (4.5,1)   (b) {$\vdots$};
\node at (.5,7)   (a) {$\vdots$};
\node at (4.5,7)   (b) {$\vdots$};
\begin{knot}[flip crossing/.list={}]
\strand (0,10)--(1,10)--(4,4)--(5,4);
\strand (0,9)--(1,9)--(4,3)--(5,3);
\strand (0,8)--(1,8)--(4,2)--(5,2);
\strand (0,6)--(1,6)--(4,0)--(5,0);
\strand (0,4)--(1,4)--(4,10)--(5,10);
\strand (0,3)--(1,3)--(4,9)--(5,9);
\strand (0,2)--(1,2)--(4,8)--(5,8);
\strand (0,0)--(1,0)--(4,6)--(5,6);
\end{knot}
\end{tikzpicture}
\captionof{figure}{$\bigswap$ gate.}
\end{subfigure}
\end{figure}
For $\bigswap$ $m^2$ swaps are needed if we arrange the qubits on a line, or $m$ swaps otherwise.
This concludes the description of our first subroutine.

To complete the description of our quantum algorithm we need to estimate the derivative of the cost function. This can be achieved by exploiting the following subroutine.
\subsection{Subroutine 2}
Subroutine $2$ implements the channel $\mathcal{E}^l$. This part of the algorithm takes as input $m_{l-1}$ qubits in the state $\rho^{l-1}$, where $m_{l-1}$ is the number of qubits in layer $l-1$. The output is $\rho^l=\mathcal{E}^l(\rho^{l-1})$.
\begin{itemize}
\item[Step 2a:]\textbf{Initialization}\\
Tensor $m_l$ qubits in state $\ket{0}$ with the input: 
\begin{equation*}
	\rho^{l-1}\rightarrow\rho^{l-1}\otimes\underbrace{\ket{0}\bra{0}\otimes\hdots\otimes\ket{0}\bra{0}}_{m_l}.
\end{equation*}
Recources: In this step $m_{l-1}+m_l$ qubits are required. \item[Step 2b:] \textbf{Perceptrons}\\
Apply the perceptrons in layer $l$:
\begin{equation*}
	\rho^{l-1}\otimes\ket{0\hdots0}\bra{0\hdots0} \rightarrow \left(\prod_{k=1}^{n_l}U_k^l\right) \rho^{l-1}\otimes\ket{0\hdots0}\bra{0\hdots0}\left(\prod_{k=1}^{n_l}U_k^l\right)^\dagger=\tilde{\rho}^{l-1,l}.
\end{equation*}
Resources: We require $m_{l-1}+m_l$ qubits and $n_l$ gates.
\item[Step 2c:]\textbf{Partial trace}\\
Take the partial trace over layer $m_{l-1}$:
\begin{equation*}
	\tilde{\rho}^{l-1,l}\xrightarrow{\tr}\rho^l.
\end{equation*}
Resources: In this step we go from $m_{l-1}+m_{l}$ qubits to $m_l$ qubits without any gates.
\end{itemize}
To get $\rho_{\text{out}}$ from $\rho_{\text{in}}$ we need to repeat Steps $2a$ to $2c$ a total of $L$ times. The total number of qubits required to carry out this subroutine is given by $\max\{m_1+m_2,m_2+m_3,\hdots,m_L+m_{\text{out}}\}$. We need to apply $n_1+n_2+\hdots+n_L$ perceptrons, where $n_i$ is the number of perceptrons in layer $i$.
\subsection{Algorithm for the cost function}
Putting it all together we can estimate the cost function via three steps:
\begin{itemize}
\item[Step 1:]Prepare $2$ copies of the state $\ket{\phi_x}$ with probability $\frac{1}{N}$.
\item[Step 2:] Do subroutine $2$ on the last $m$ qubits.
\item[Step 3:] Do $\swap$ trick. 
\item[Step 4:] Repeat Steps 1,2, and 3 a total of $M$ times for same value of $x$ to estimate $\braket{\phi_x|\rho|\phi_x}$. (The choice of $M$ affects the accuracy of the latter; the bigger $M$ the more accurate we get.) \end{itemize}
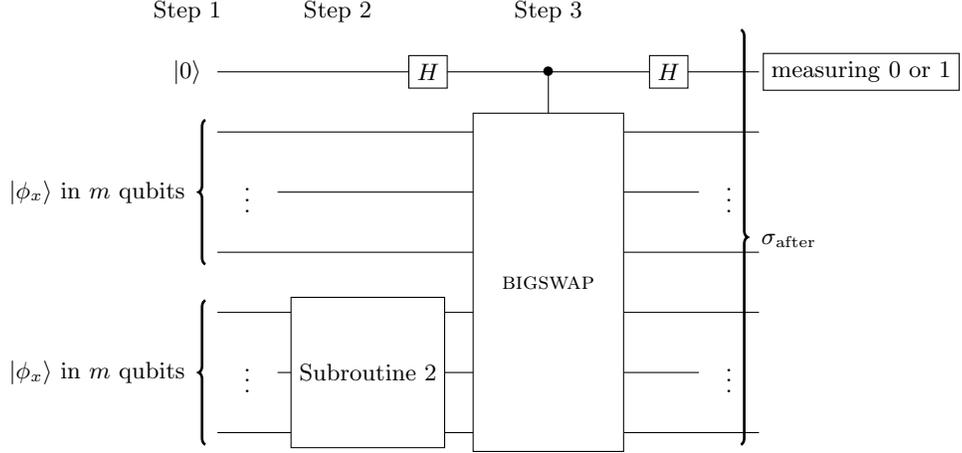
\begin{figure}[H]
\centering
\begin{tikzpicture}[scale=.8, decoration={brace}]
\draw(-2,6)--(7,6);	
\draw(-2,5)--(7,5);	
\draw(-1,4)--(6,4);	
\draw(-2,3)--(7,3);	
\draw(-2,2)--(7,2);	
\draw(-1,1)--(6,1);	
\draw(-2,0)--(7,0);	
\draw (3.5,2.5)--(3.5,6);
\node (a) at (3.5,6) {\textbullet};
\node (a) at (-1.5,1) {$\vdots$};
\node (b) at (6.5,1) {$\vdots$};
\node (c) at (-1.5,4) {$\vdots$};
\node (d) at (6.5,4) {$\vdots$};
\node[draw,minimum height=4.5cm,minimum width=2cm,fill=white] (e) at (3.5,2.5) {$\bigswap$};
\node[draw,fill=white] (f) at (1.5,6) {$H$};
\node[draw,fill=white] (g) at (5.5,6) {$H$};
\node[draw,fill=white] (h) at (8.7,6) {measuring 0 or 1};
\node (i) at (-2.5,6) {$\ket{0}$};
\node (i) at (-4.0,4) {$\ket{\phi_x}$ in $m$ qubits};
\node (e) at (-4.0,1) {$\ket{\phi_x}$ in $m$ qubits};
\node (q) at (7.5,3.2) {$\sigma_{\text{after}}$};
\draw [decorate,line width=1pt] (-2.2,-.2) -- (-2.2,2.2);
\draw [decorate,line width=1pt] (-2.2,2.8) -- (-2.2,5.2);
\draw [decorate,line width=1pt] (6.7,6.7) -- (6.7,-.2);
\node[draw,minimum height=2cm,minimum width=2cm,fill=white] (e) at (.5,1) {Subroutine 2};
\node (k) at (-2.5,7) {Step 1};
\node (u) at (0,7) {Step 2};
\node (y) at (3.5,7) {Step 3};
\end{tikzpicture}
\captionof{figure}{Steps 1 to 3 of Algorithm.}
\end{figure}

Choose $x$ randomly $N$ times and employ this algorithm each time to compute the expectation value over $x$ and thus the cost function $C=\frac{1}{N}\sum_x \braket{\phi_x|\rho|\phi_x}$. The total number of gates and perceptrons required is $N\times M (\sum_{i=1}^L n_i + 3)$ . The number of qubits required is $\leq 2\times W+m+1$, where $W$ is the width of the QNN, i.e., $W=\max\{m_1,\hdots,m_{\text{out}}\}$.

\subsection{Algorithm for derivative}
To work out the derivative $\frac{dC}{ds}$ of the cost function we compute $\frac{\delta C}{\delta x^\alpha}$, where $x^\alpha$ is the vector of all the parameters. For a single three-qubit perceptron $U=e^{ik}$ with $k=\sum k_{\alpha,\beta,\gamma}\sigma^\alpha\otimes \sigma^\beta \otimes \sigma^\gamma$ we write
\begin{equation*}
\left.x^\alpha=\begin{pmatrix}k_{000}\\k_{001}\\k_{002}\\k_{003}\\k_{010}\\\vdots\\\end{pmatrix}\right\} = 64 \text{ parameters.}
\end{equation*}
For a four-qubit QNN with two three-qubit perceptrons, see Figure \ref{fig:22QNN}, we have
\begin{equation}
\left.x^\alpha=\begin{pmatrix}k^1_{000}\\\vdots\\k^1_{333}\\k^2_{000}\\\vdots\\k^2_{333}\end{pmatrix}\right\} = 2\times64 \text{ parameters.} 
\end{equation}

Now we need to work out $\frac{\delta C}{\delta X^\alpha}\approx\frac{C(x+\epsilon^\alpha)-C(x)}{\epsilon}$, where 
$$\epsilon^\alpha=\begin{pmatrix}0\\\vdots\\0\\\epsilon\\0\\\vdots\\0\end{pmatrix},$$ 
i.e. $\epsilon$ is the $\alpha$th entry and $\alpha=1,\hdots,(\# \text{perc})\times64$.

Suppose we know $C(x)$ and work out $\frac{\delta C}{\delta X^\alpha}$, $Q=(\# \text{perc})\times64$ times. This gives us
$$x=
	\begin{pmatrix}
\frac{\delta C}{\delta X^1}\\
\frac{\delta C}{\delta X^2}\\
\vdots\\
\frac{\delta C}{\delta X^Q}
\end{pmatrix}.
$$
All that is left to do is the gradient ascent step, with
$$x_{\text{new}}=x_{\text{old}}+\frac{1}{2\lambda}\begin{pmatrix}
\frac{\delta C}{\delta X^1}\\
\frac{\delta C}{\delta X^2}\\
\vdots\\
\frac{\delta C}{\delta X^Q}
\end{pmatrix}.
$$
This always makes the cost function larger: 
\begin{align*}
C(x_\text{new})&=C(x_\text{old}+\frac{1}{2\lambda}\frac{\delta C}{\delta x})\\
 &\approx C(x_\text{old})+\sum_{\alpha=1}^Q \frac{1}{2\lambda}\frac{\delta C}{\delta X^\alpha} \frac{\delta C}{\delta X^\alpha}\\
C(x_\text{new})-C(x_\text{old})&\approx\sum(\text{positive}).
\end{align*}

\bibliographystyle{apsrev4-1}
\bibliography{paper}
\end{document}